\documentclass[11pt]{article}
\usepackage{amsmath,amssymb,latexsym}
\usepackage[T1]{fontenc}
\usepackage{epsfig}
\usepackage{graphicx}
\usepackage{makeidx}
\usepackage{slashed}
\usepackage{color}
\usepackage{cite}
\usepackage{subcaption}
\usepackage{feynmf}
\long\def\symbolfootnote[#1]#2{\begingroup%
\def\thefootnote{\fnsymbol{footnote}}\footnote[#1]{#2}\endgroup}

 \newlength{\dinwidth} \newlength{\dinmargin}
\begin{document}

\begin{flushright}
FR-PHENO-2019-008\\
Nikhef/2019-010
\end{flushright}

\vspace*{1.5cm}

\begin{center}
{\Large\sc Next-to-leading power threshold effects for resummed prompt photon production}\\[10ex] 
 \vspace*{1cm} 
 Melissa van Beekveld$^{a,b}$,
 Wim Beenakker$^{a,c}$, 
 Rahul Basu$^{d}\symbolfootnote[2]{Deceased. We dedicate this paper to his memory.}$, 
 Eric Laenen$^{b,c,e}$, 
 Anuradha Misra$^f$, 
 Patrick Motylinski$^g$\\[1cm]
 {\it 
$^a$ {Theoretical High Energy Physics, Radboud University
Nijmegen, Heyendaalseweg 135, 6525~AJ Nijmegen, The Netherlands}\\
$^b${Nikhef, Science Park 105, 1098 XG Amsterdam, The Netherlands}\\
$^c$ {Institute of Physics, University of Amsterdam, Science Park 904, 1018 XE Amsterdam, The Netherlands}\\
$^d$ {The Institute of Mathematical Sciences, CIT Campus,
 Taramani, Chennai 600 113, India}\\
 $^e${ITF, Utrecht University, Leuvenlaan 4, 3584 CE Utrecht, NL}\\
 $^f$ {Department of Physics, University of Mumbai, Santacruz(E)
  Mumbai, 400\,098, India}\\
$^g$ {Physikalisches Institut, Albert-Ludwigs-Universit\"{a}t Freiburg,
Hermann-Herder-Stra{\ss}e 3, D-79104 Freiburg i.Br., Germany}}\\
\end{center}

\begin{abstract}
\noindent{
We assess and compare different methods for  including leading threshold logarithms at next-to-leading-power
in prompt photon production at hadron colliders, for both the
direct and parton fragmentation mechanisms. 
We do this in addition to next-to-leading logarithmic threshold and joint resummation at leading power. 
We study the size of these effects and their scale variations for LHC kinematics. We find that the next-to-leading power effects  have a noticeable effect on the photon transverse momentum distribution, typically of order $\mathcal{O}(10\%)$, depending on the method of inclusion. Our results indicate that next-to-leading power terms can reduce the scale dependence of the distribution considerably.}
\end{abstract}

\vspace*{\fill}

\newpage
\reversemarginpar

\parindent=0pt

\section{Introduction}
\label{sec:introduction}
We consider the production of a prompt photon with a given transverse momentum $p_T$ in hadronic collisions. This production can either proceed directly or through fragmentation of a parton. The corresponding perturbative QCD description can involve sizable corrections from soft and collinear parton emissions. These follow from the presence of a 
threshold at $S=4{p_T}^2$ and take the form of
large logarithmic corrections
\cite{Sterman:1987aj,Catani:1989ne}. This is a general property of hadronic cross-sections differential in a variable $\xi$ expressing the distance to
threshold, which may be written schematically as
\begin{equation}
  \frac{d \sigma}{d \xi} \, \sim \, 
  \sum_{n = 0}^{\infty} \left( \frac{\alpha_s}{\pi}
  \right)^n \sum_{m = 0}^{2 n - 1} \left[ \, c_{n m}^{(-1)} \left(
      \frac{\ln^m \xi}{\xi} \right)_+ + \, c_{n}^{(\delta)} \,
    \delta(\xi) + \, c_{nm}^{(0)} \, \ln^m \xi + \ldots \, \right] .
  \label{thresholddef}
\end{equation}
The \emph{first term} in \eqref{thresholddef} gathers all plus distributions in $\xi$.
The all-order resummation of such terms has been extensively researched,
leading to a variety of successful resummation approaches. In the diagrammatic approach, which we use here, the set of diagrams most relevant for the cross-section near
threshold is factorized into hard, soft and jet functions. All-order expressions are then constructed for each \cite{Sterman:1987aj,Catani:1989ne}.  For direct prompt photon production this led to threshold resummation 
 at next-to-leading logarithmic (NLL) accuracy \cite{Laenen:1998qw,
  Catani:1998tm,
  Catani:1999hs,
  Kidonakis:1999hq, Sterman:2000pt,
  Sterman:2004yk},  including partial results at
higher order derived from resummation.
This was extended to include production via fragmentation
\cite{deFlorian:2005wf,
  deFlorian:2005yj, deFlorian:2013taa}.
A somewhat different approach based on the renormalization group was taken in \cite{Bolzoni:2005xn}.
Prompt photon threshold resummation was also performed in the context
of soft-collinear effective theory (SCET) \cite{Becher:2009th,Becher:2012xr},
where the hard, soft and jet
functions are each described by their own effective theory, matched together appropriately; resummation then follows
from renormalization group evolution and matching conditions among
these functions. Results at next-to-next-to-leading logarithmic (NNLL) accuracy for single-particle inclusive kinematics were obtained in \cite{Becher:2009th,Becher:2012xr,Hinderer:2018nkb}, and at full N$^3$LL accuracy in \cite{Schwartz:2016olw}.

In \cite{Laenen:2000ij,Li:1998is}
it was shown, to NLL accuracy,  that threshold logarithms can be
resummed jointly with recoil corrections. This joint resummation has been applied to
direct prompt photon production in \cite{Laenen:2000de}, heavy quark production in \cite{Banfi:2004xa}, BSM processes in \cite{Bozzi:2007tea,Fuks:2007gk,Debove:2011xj} and to 
vector boson and Higgs production in
\cite{Kulesza:2002rh,Kulesza:2003wn}. The latter references 
 introduced resummation via PDF evolution, which we use in the present paper as well.
Joint resummation for vector boson production was extended to NNLL
accuracy in \cite{Marzani:2016smx}, and was improved in \cite{Muselli:2017bad} to match better to single resummation results. Joint resummation beyond NLL accuracy was also achieved in the SCET framework \cite{Lustermans:2016nvk}.

The {\it second term} in \eqref{thresholddef} corresponds to virtual
contributions, the exact knowledge of which is important for exact fixed order calculations, which are now known to next-to-next-to-leading order (NNLO) \cite{Campbell:2016lzl}. The {\it third term}
contains all integrable logarithmic effects, which are suppressed by
one power of $\xi$ with respect to the first term. These
{\it next-to-leading power} (NLP) corrections have been the subject of much recent
research, and are the focus of the present study for the case of prompt photon production.
Logarithmic NLP terms have been shown to be numerically significant in certain scattering processes~\cite{Kramer:1996iq,Herzog:2014wja}. The structure of such subleading soft corrections was first clarified in 
refs.~\cite{Low:1958sn,Burnett:1967km}, with ref.~\cite{DelDuca:1990gz}
extending this analysis to massless particles. Using path-integral
methods ref.~\cite{Laenen:2008gt} constructed effective Feynman
rules for next-to-soft emissions and showed that a large class of NLP contributions exponentiates, specifically those arising from emissions off external lines. This was confirmed in a diagrammatic analysis~\cite{Laenen:2010uz}.
Another approach, using physical evolution kernels, achieved  the resummation of logarithmic NLP corrections for Drell-Yan and Higgs production
\cite{Soar:2009yh,Moch:2009hr,Moch:2009mu,deFlorian:2014vta,Presti:2014lqa}. 

At leading power, resummation is intimately related to 
factorization of soft and collinear
divergences ~\cite{Contopanagos:1997nh}. In ~\cite{Bonocore:2015esa,Bonocore:2016awd} a NLP factorization formula  based on \cite{DelDuca:1990gz} was shown to have considerable predictive power among NLP logarithms. A first principles NLP factorization analysis in the context of Yukawa theory was undertaken in
\cite{Gervais:2017yxv,Gervais:2017zky,Gervais:2017zdb}.
Related analyses have been carried out in
the SCET framework~\cite{Larkoski:2014bxa,Kolodrubetz:2016uim,Moult:2016fqy,Moult:2017rpl,Feige:2017zci,Chang:2017atu,Beneke:2004in,Moult:2017jsg,Ebert:2018lzn,Ebert:2018gsn,Moult:2018jjd,Beneke:2018gvs,Bhattacharya:2018vph, Moult:2019mog} and results using either diagrammatic or effective
theory methods have been shown to be potentially useful for improving
the accuracy of fixed-order
calculations~\cite{DelDuca:2017twk,Bonocore:2014wua,Bahjat-Abbas:2018hpv,Boughezal:2016zws,Boughezal:2018mvf}. Recently,
the SCET framework has been used to demonstrate that the
leading-logarithmic (LL) NLP contributions to Drell-Yan production can
indeed be resummed~\cite{Beneke:2018gvs}.

Preliminary studies \cite{Basu:2007nu,Basu:2012ma,Misra:2018nsy} were performed
for the resummation of a large class of leading logarithmic (LL with $m=2n-1$) NLP terms for \emph{direct}
production of prompt photons, in both threshold and joint
resummation. In this paper we extend this work in a number of new
directions.  First, we now include the fragmentation mechanism,
which requires additional colour structures in the hard
scattering. Second, we assess different approaches to include initial and final state NLP terms and analyze them for both threshold and joint
resummation. We furthermore examine the effect of including the NLP terms
on the scale dependence of the resummed cross-section. Finally, we use
the prompt photon process as a case study of NLP effects 
for a final state containing color-charged particles, 
providing numerical studies to assess the size of the NLP terms.

The paper is organized as follows. In section \ref{sec:resummation} we review the
threshold and joint resummed photon $p_T$ distribution, and discuss
the inclusion of NLP effects for the initial and final state.  In
section \ref{sec:numerical-studies} we assess the numerical impact of these corrections, and we
conclude in section \ref{sec:conclusions}. In appendix \ref{app:exponents} we collect explicit expressions for quantities listed in section \ref{sec:resummation}, while in appendix \ref{app:NLOcomparison} we compare the NLO expansion of our resummed expressions at NLP with exact results.

\section{Resummation}
\label{sec:resummation}

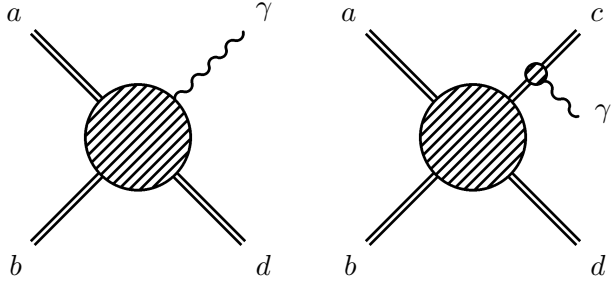
\begin{figure}
    \centering
\begin{fmffile}{kin1}
\begin{fmfchar*}(80,80)
  \fmfleft{i1,i2}
  \fmfright{f1,f2}
  \fmfforce{(0,0)}{i2} 
  \fmfforce{(0,h)}{i1} 
  \fmfforce{(.5w,.5h)}{v1}
  \fmfforce{(w,h)}{f2} 
  \fmfforce{(w,0)}{f1} 
  \fmf{double}{i1,v1}
  \fmflabel{$a$}{i1}
  \fmflabel{$b$}{i2}
  \fmflabel{$d$}{f1}
  \fmflabel{$\gamma$}{f2}
  \fmf{double}{v1,i2}
  \fmf{double}{f1,v1}
  \fmf{photon}{f2,v1}
  \fmfv{decor.shape=circle,decor.filled=shaded, d.si=.5w}{v1}
\end{fmfchar*}
\end{fmffile}
\hspace{1cm}
\begin{fmffile}{kin2}
\begin{fmfchar*}(80,80)
  \fmfleft{i1,i2}
  \fmfright{f1,f2,f3}
  \fmfforce{(0,0)}{i2} 
  \fmfforce{(0,h)}{i1} 
  \fmfforce{(.5w,.5h)}{v1}
  \fmfforce{(.8w,.8h)}{v3}
  \fmfforce{(.8w,.8h)}{v2}
  \fmfforce{(w,h)}{f3} 
  \fmfforce{(w,.6h)}{f2} 
  \fmfforce{(w,0)}{f1} 
  \fmf{double}{i1,v1}
  \fmflabel{$a$}{i1}
  \fmflabel{$b$}{i2}
  \fmflabel{$d$}{f1}
  \fmflabel{$\gamma$}{f2}
  \fmflabel{$c$}{f3}
  \fmf{double}{v1,i2}
  \fmf{double}{v1,f1}
  \fmf{double}{v1,f3}
  \fmf{photon}{f2,v2}
  \fmfv{decor.shape=circle,decor.filled=shaded, d.si=.5w}{v1}
  \fmfv{decor.shape=circle,decor.filled=empty, d.si=.1w}{v3}
  \fmfv{decor.shape=circle,d.si=.1w,decor.filled=shaded}{v2}
\end{fmfchar*}
\end{fmffile}
\vspace{0.5cm}
    \caption{Diagrams for the direct production of a photon (left) and for the production of a photon via fragmentation (right), where the double lines indicate color charged partons. }
    \label{fig:process}
\end{figure}

We consider the inclusive transverse momentum distribution of photons produced
at fixed $p_T$ in proton-proton collisions 
\begin{equation}
  \label{eq:el-proc}
  h_A(p_A) + h_B(p_B) \to \gamma(p_\gamma)+X \,,
\end{equation}
where $h_{A,B}$ refers to the two incoming protons
 and $X$ to the unobserved part of the final state. 
The lowest order QCD processes producing the photon directly
at partonic center-of-mass (CM) energy $\sqrt{s}$ are
\begin{equation}
  \label{eq:parton-proc}
  \begin{split}
    &q(p_a) + \bar q(p_b) \to \gamma(p_\gamma) + g(p_d) \\
    &g(p_a) +  q(p_b) \to \gamma(p_\gamma) + q(p_d)\,,
  \end{split}
\end{equation}
where in the second reaction $q$ stands for both quark and anti-quark. The partonic momenta $p_a$ and $p_b$ are related to the hadronic ones via $p_a = x_a p_A$ and $p_b = x_b p_B$. In the CM frame of the initial state particles it is convenient to parametrize the photon momentum $p_{\gamma}$ as
\begin{eqnarray*}
p_{\gamma} = \Big(p_T\cosh(\eta),{\bf p}_T, p_T\sinh(\eta)\Big),
\end{eqnarray*}
where ${\bf p}_T$ and $\eta$ are the transverse momentum and pseudorapidity of the photon, 
and we denote $|{\bf p}_T|$ by $p_T$. 
The minimum value of ${\rm cosh}(\eta)$ is equal to $ 1$, therefore the partonic threshold is at $s = 4p_T^2$. The distance $\xi$ to threshold in \eqref{thresholddef} can be written as $\xi=1-\hat{x}_T^2$, where $\hat{x}_T = \frac{2p_T}{\sqrt{s}}$. The hadronic equivalent of $\hat{x}_T$ is denoted as $x_T = \frac{2p_T}{\sqrt{S}}$.
Apart from the partonic processes that produce the photon directly there are contributions from $2 \to 2$ parton scattering
\begin{equation}
  \label{eq:frag-proc}
  a(p_a) + b(p_b) \to c(p_c) + d(p_d)\,, 
\end{equation}
where the photon is subsequently produced by fragmentation of 
final state parton $c$. In this paper 
the fragmenting parton will be either a quark or anti-quark (since gluons only couple to photons at one order higher in $\alpha_s$). Fig.~\ref{fig:process} schematically illustrates the direct and fragmentation production
processes.
Note that the fragmentation component 
contributes at $\mathcal{O}(\alpha \alpha_s)$. Although it is sub-dominant, i.e. the fragmentation function behaves as $1/N$ \cite{Catani:1999hs}, threshold resummation can substantially 
enhance this component \cite{deFlorian:2005wf}, so that it must be taken into account in this analysis. 
The combined differential cross-section for prompt photon production is therefore a sum of two parts 
\begin{eqnarray}
  \label{eq:1}
 p_T^3 \frac{ {\rm d} \sigma^{({\rm comb})}_{AB\to \gamma+X}}{{\rm d}p_T}
  &=&  p_T^3\frac{ {\rm d} \sigma^{({\rm direct})}_{AB\to \gamma+X}}{{\rm d}p_T}
  +  p_T^3\frac{ {\rm d} \sigma^{({\rm frag})}_{AB\to \gamma+X}}{{\rm d}p_T},
\end{eqnarray}
where the two terms correspond to the subprocesses \eqref{eq:parton-proc}
and \eqref{eq:frag-proc}. Note that we have rescaled both terms with powers of $p_T$ to make them dimensionless.\\

In what follows we assume that any radiation additional to the Born process modifies the final state only slightly, because it will be soft and/or collinear to initial or final state partons. In higher order corrections,
powers of the threshold logarithm $\ln(1-\hat{x}_T^2)$ will appear. These can 
be resummed in the framework of either threshold or joint resummation, both of which we now review.

\subsection{Threshold resummation}
\label{sec:thresholdresummation}

The $p_T$ distribution for the direct production of a photon in hadronic collisions may be described in the context of collinear factorization
\begin{eqnarray}
\label{equation:hadronicdist}
p_T^3 \frac{{\rm d} \sigma^{(\mathrm{direct})}_{AB\to \gamma + X}(x_T^2)}{{\rm d}p_T}
&=&  \sum_{a,b}\  \int_{0}^1 {\rm d} x_a \hspace{1mm}f_{a/A}(x_a,\mu_F) \int_{0}^1 {\rm d} x_b \hspace{1mm}f_{b/B}(x_b,\mu_F) \nonumber \\
&&\hspace{2cm} \times \int_0^1{\rm d} \hat{x}_T^2 \hspace{1mm} \delta\left(\hat{x}_T^2 - \frac{x_T^2}{x_ax_b}\right) p_T^3 \frac{{\rm d}\sigma_{ab\rightarrow \gamma d}(\hat{x}_T^2,\mu,\mu_F)}{{\rm d} p_T},
\end{eqnarray} 
where the sum is over parton flavours $a,b$ and the parton distribution functions (PDFs) are indicated by $f_{i/I}(x_i,\mu_F)$. The scale $\mu_F$ ($\mu$) denotes the factorization (renormalization) scale. 
The partonic differential cross-section ${\rm d}\sigma/{\rm d}p_T$ has a perturbative expansion in $\alpha_s$. Near threshold, this expansion can be written as a $2\rightarrow 2$ hard scattering subprocess dressed with additional soft and collinear radiation. At leading order, the $p_T$ distribution for the $2\rightarrow 2$ partonic process is given by
\begin{eqnarray*}
p_T^3 \frac{{\rm d}\sigma^{({\rm direct})}_{ab\to \gamma d}}{{\rm d}p_T}
&=& \frac{p_T^4}{8\pi s^2}\frac{|\mathcal{M}_{ab\to\gamma d}(\hat{x}_T^2)|^2}{\sqrt{1-\hat{x}_T^2}}\,,
\end{eqnarray*}
with $|\mathcal{M}_{ab\to\gamma d}(\hat{x}_T^2)|^2$ the color- and spin-averaged square of the lowest order amplitude. 
The effects of additional soft gluon radiation dominate the perturbative corrections in the regime where $\hat{x}_T^2 \rightarrow 1$ and the resulting logarithms must be resummed to all orders in $\alpha_s$ for the perturbative series to remain useful. 
It is convenient for us to work in Mellin space
\begin{eqnarray}
\label{eq:hadronic}
p_T^3\frac{{\rm d}\sigma^{\rm (direct)}_{AB\to \gamma + X}(N)}{{\rm d}p_T} &\equiv& \int_0^1{\rm d}x_T^2 \,(x_T^2)^{N-1} \frac{p_T^3 {\rm d} \sigma^{({\rm direct})}_{AB\to \gamma + X}(x_T^2)}{{\rm d} p_T} \nonumber \\
&=& \sum_{a,b} \int {\rm d}x_a \,x_a^{N} f_{a/A}(x_a,\mu_F)\int {\rm d}x_b \,x_b^{N}f_{b/B}(x_b,\mu_F)\nonumber \\
&&\hspace{1cm} \times \int_0^1{\rm d}\hat{x}_T^2 \,(\hat{x}_T^2)^{N-1} p_T^3\frac{{\rm d}\sigma_{ab\to \gamma d}(\hat{x}_T^2)}{{\rm d}p_T}\nonumber \\
&\equiv&\sum_{a,b}f_{a/A}(N+1,\mu_F)f_{b/B}(N+1,\mu_F) \; p_T^3\frac{{\rm d}\sigma^{\rm (direct)}_{ab\to \gamma d}(N)}{{\rm d}p_T}\,,
\end{eqnarray}
where we have suppressed the scale dependence in the partonic cross-section.
The threshold limit $\hat{x}_T^2 \rightarrow 1$ corresponds to the Mellin moment limit $N\rightarrow \infty$. To {\it leading power} in the threshold variable, one can factorize the additional soft and collinear radiation from the hard part of the scattering cross-section into soft and jet functions \cite{Laenen:1998qw,Catani:1998tm,Catani:1999hs,Kidonakis:1999hq, Sterman:2004yk}. Resummation organizes the contributions of these
into exponential form, and leads to the expression
\begin{eqnarray}
p_T^3\frac{{\rm d}\sigma^{\rm (direct,thres)}_{ab\to \gamma d}(N)}{{\rm d}p_T} &=&  \int_0^1 {\rm d}\hat{x}_T^2 (\hat{x}_T^2)^{N-1} \frac{p_T^4}{8\pi s^2}\frac{|\mathcal{M}_{ab\to\gamma d}(\hat{x}_T^2)|^2}{\sqrt{1-\hat{x}_T^2}}C_{\delta}^{(ab\to\gamma d)}(\alpha_s,\hat{x}_T^2) \nonumber\\[0.4ex]
&& \hspace{-4cm}\times \exp\left[E_a^{\rm PT}(N,Q,\mu_F,\mu)+E_b^{\rm PT}(N,Q,\mu_F,\mu) + F_d (N,Q,\mu)+ g_{abd} (N,\mu)\right]\nonumber \\[0.8ex]
 &&\hspace{-4cm}\equiv  \int_0^1 {\rm d}\hat{x}_T^2 (\hat{x}_T^2)^{N-1} \frac{p_T^4}{8\pi s^2}\frac{|\mathcal{M}_{ab\to\gamma d}(\hat{x}_T^2)|^2}{\sqrt{1-\hat{x}_T^2}}C_{\delta}^{(ab\to\gamma d)}(\alpha_s,\hat{x}_T^2)
 \, P_{abd}(N,Q,\mu_F,\mu)\,,
\end{eqnarray} 
where $C^{(ab \to \gamma d)}_{\delta} \equiv 1 + (\alpha_s/\pi)C^{(1)}_{ab \to \gamma d} + \mathcal{O}(\alpha_s^2) $ is the perturbative hard part of the scattering process. The large logarithmic corrections are organized in a function $P_{abd}(N,Q,\mu_F,\mu)$, whose definition is shown on the second line. The initial state contributions are indicated by $E^{\rm PT}_{a,b}$, the final state jet contribution by $F_d$, and soft wide angle 
radiation is summarized by $g_{abd}$. Expressions for these functions are given in appendix \ref{app:exponents}. The hard scale of the process is denoted by $Q$, which will be chosen to be $2p_T$ for our numerical studies.\\
The hadronic resummed $p_T$ distribution now follows from substituting the expression for the resummed partonic distribution into \eqref{eq:hadronic} and taking an inverse Mellin transform, resulting in
\begin{eqnarray}
  \label{eq:thresholddirect}
   p_T^3  \frac{{\rm d} \sigma^{({\rm direct, thres})}_{AB\to \gamma + X}(x_T^2)}{{\rm d}p_T}
&=&\frac{p_T^4}{8\pi S^2} \sum_{a,b}\  \int_{\cal C} \frac{{\rm d}N}{2 \pi i}
(x_T^2)^{-N-1}f_{a/A}(N,\mu_F) f_{b/B}(N,\mu_F)\nonumber \ \\
&& \hspace{-2cm} \times \int_0^1 {\rm d}\hat{x}_T^2 (\hat{x}_T^2)^{N} \frac{|\mathcal{M}_{ab\to\gamma d}(\hat{x}_T^2)|^2}{\sqrt{1-\hat{x}_T^2}}C_{\delta}^{(ab\to\gamma d)}
\left(\alpha_s,\hat{x}_T^2\right)\,P_{abd}(N,Q,\mu_F,\mu)\,,
\end{eqnarray}
where we have used
\begin{eqnarray*}
\frac{p_T^4}{s^2} = \frac{p_T^4}{S^2}\,\hat{x}_T^2\, \frac{1}{x_a x_b}\,\frac{1}{x_T^2}\,.
\end{eqnarray*}
The inverse Mellin transform must formally be taken over a  straight line that runs from $c-i\infty$ to $c+i\infty$ with $c$ chosen such that it runs to the right of all singularities\footnote{In practice we choose it according to the minimal prescription (MP)\cite{Catani:1996yz}.}. For better numerical convergence, it is useful to pick a contour  $\mathcal{C}=C_{{\rm MP}}+y\,{\rm e}^{i\phi}$, where $C_{{\rm MP}} > 0$ is a real constant, $\phi$ is the angle with respect to the real $N$-axis and $y$ runs from $-\infty$ to $0$ for $\phi=-\phi_{\rm MP}$ and from $0$
to $\infty$ for $\phi = \phi_{\rm MP}$,
with $\phi_{\rm MP} > \pi/2$ \cite{Vogt:2004ns}. 

The threshold-resummed $p_T$ distribution for the fragmentation component can be derived \cite{deFlorian:2005wf} in analogy to the direct component and reads
\begin{eqnarray}
  \label{eq:thresholdfrag}
    p_T^3\frac{{\rm d} \sigma^{({\rm frag, thres})}_{AB\to \gamma + X} (x_T^2)}{{\rm d}p_T}
&=&\frac{p_T^4}{8\pi S^2} \sum_{a,b,c}\  \int_{\cal C} \frac{{\rm d}N}{2 \pi i}
(x_T^2)^{-N-1}f_{a/A}(N,\mu_F) f_{b/B}(N,\mu_F)D_{\gamma/c}(2N+1,\mu_F)  \nonumber \\
&& \hspace{-0.9cm}\times \int_0^1 {\rm d}\hat{x}_T^2\, (\hat{x}_T^2)^{N} \,\frac{|\mathcal{M}_{ab\to c d}(\hat{x}_T^2)|^2}{\sqrt{1-\hat{x}_T^2}}\,C_{\delta}^{(ab\to c d)}(\alpha_s,\hat{x}_T^2)\hspace{0.1cm} P_{abcd}(N,Q,\mu_F,\mu)\,.
\end{eqnarray}
There are two extra ingredients in this expression with respect to direct photon production. First, the $N$-space fragmentation function  $D_{\gamma/c}(2N+1,\mu_{F})$ is included, corresponding to the probability of parton $c$ fragmenting into a photon, where we have chosen the fragmentation scale equal to the factorization scale $\mu_F$. Second, there are now a larger number of color structures that can connect the external
partons in this process. This results in a soft anomalous dimension {\em matrix} $\Gamma_s$ governing the soft wide angle emission. In a color basis in which $\Gamma_s$ is diagonal \cite{Kidonakis:1999hq} the profile function $P_{abcd}$ is given by \cite{deFlorian:2005wf,Basu:2012ma}
\begin{align}
 \label{eq:11}
 & P_{abcd}(N,Q,\mu_F,\mu)  = \exp\Bigg[ E^{\rm PT}_a(N,Q, \mu_F, \mu)+E^{\rm PT}_b (N,Q,\mu_F, \mu)\nonumber \\
&\hspace{5mm}  +E^{\rm PT}_c(N,Q,\mu_{F}, \mu)+F_d(N,Q,\mu)\Bigg] \times \left[\sum_I G^I_{ab \to cd}
  \exp\left(\Gamma^{I,\mathrm{(int)}}_{ab \to cd}(N) \right) \right]\,.
\end{align}
The sum on the last line runs over all possible color configurations $I$ with $G^I_{ab \to cd} $ representing a relative weight for each color configuration such that $\sum_I G^I_{ab \to cd} = 1$.
The associated soft anomalous dimensions $\Gamma^{I,(\mathrm{int})}_{ab \to cd}(N)$  are given by 
\begin{equation} 
\Gamma^{I,\mathrm{(int)}}_{ab \to cd}(N)= \int_0^1 {\rm d}z\, \frac{z^{N-1}-1}{1-z} D_{I,ab \to cd}\left(\alpha_s\left (\left(1-z\right)^2Q^2\right)\right)\,,
\end{equation}
which to NLL accuracy is equal to
\begin{equation}
\Gamma^{I,\mathrm{(int)}}_{ab \to cd}(N) = \frac{D_{I,ab \to cd}^{(1)}}{2\pi b_0}\ln(1-2\lambda) + O\left(\alpha_s(\alpha_s \ln N)^k \right)\,.
\end{equation}
Here $b_0$ is the first term of the beta function for the strong coupling (see appendix \ref{app:exponents}) and $\lambda = b_0 \alpha_s \ln (N e^{\gamma_E})$. The coefficients $D_{I,ab \to cd}^{(1)}$, the color weights $G_{ab\to cd}^I$ , the one loop hard matching coefficients $C^{(1)}_{ab \to cd}$, and the $N$-space expression for the $p_T$ distribution of the partonic subprocess $ab\rightarrow cd$ are given in the appendix of
\cite{deFlorian:2005yj}. Note that we use a Mellin transform of the partonic $p_T $ distribution shifted by $N\rightarrow N-1$ with respect to the expressions in \cite{deFlorian:2005yj}.

\subsection{Joint resummation}

Joint resummation takes into account the recoil of the hard scattering {\em process} against additional radiated partons with collective transverse momentum ${\bf Q}_T$.  This implies that the photon transverse momentum to be produced by the hard scattering is only ${\bf p}'_T = {\bf p}_T - {\bf Q}_T/2$, which effectively lowers the partonic threshold. The corresponding new scaling variable is $\tilde{x}_T^2 = 4{p'_T}^{2}/Q^2$, where $Q$ is the invariant mass of the photon-parton pair in the recoil frame. The variables $x_T^2$ and $\tilde{x}_T^2$ are related by
\begin{eqnarray*}
\tilde{x}_T^2 = x_T^2\left(\frac{S}{Q^2}\frac{{p'_T}^{2}}{p_T^2}\right)\,.
\end{eqnarray*}
It was shown in \cite{Laenen:2000de,Laenen:2000ij} that resummation of threshold and recoil logarithms can be jointly performed for {\it{sufficiently small}} values of $|{\bf Q}_T|\equiv Q_T$. 
The expression for the joint-resummed $p_T$ distribution of the direct component reads\footnote{A factor $p'_T dp'_T/dp_T$ is also present but may be put equal to $p_T$ up to a term of $\mathcal{O}(Q_T)$. This latter term is non-singular, but may contribute at NLP order beyond leading logarithm. Since a general resummation framework for these terms is yet to be developed, we neglect this contribution in the present study.}  \cite{Laenen:2000ij,Basu:2007nu,Basu:2012ma}
\begin{eqnarray}
  \label{eq:direct}
  p_T^3   \frac{{\rm d} \sigma^{({\rm direct,joint})}_{AB\to \gamma + X} (x_T^2)}{{\rm d}p_T}
&=& \frac{p_T^4}{8 \pi S^2}\ \sum_{a,b}\  \int_{\cal C} \frac{{\rm d}N}{2 \pi i}  f_{a/A}(N,\mu_F) f_{b/B}(N,\mu_F) \nonumber \\[1ex]
&& \hspace{-2cm} \times \int \frac{{\rm d}^2 {\bf Q}_T}{(2\pi)^2}\;
\left( \frac{S}{4 |{\bf p}_T - {\bf Q}_T/2|^2} \right)^{N+1}\int_0^1 {\rm d}\tilde{x}_T^2 (\tilde{x}_T^2)^{N} \frac{|\mathcal{M}_{ab\to\gamma d}(\tilde{x}_T^2)|^2}{\sqrt{1-\tilde{x}_T^2}}C_{\delta}^{(ab\to\gamma d)}(\alpha_s,\tilde{x}_T^2) \nonumber \\
&\ & \hspace{-2cm} \times \;
\int {\rm d}^2 {\bf b}\; {\rm e}^{i {\bf b} \cdot {\bf Q}_T} \;
\theta\left(\bar{\mu}-|{\bf Q}_T|\right) P_{abd}(N,b,Q,\mu_F,\mu).
\end{eqnarray}
Singular $Q_T$ behavior is readily organized using an impact parameter $\mathbf{b}$, hence the Fourier transform in the third line of this equation.
The kinematic factor linking the threshold and recoil effects
\begin{equation}
\label{eq:kinfactor}
\left(\frac{S}{4|{\bf p}_T-{\bf Q}_T/2|^2}\right)^{N+1}\,,
\end{equation} 
arises due to the inverse Mellin transform over $x_T^2$. Finally, the exponential factors are included in the "profile" function
$P_{abd}$ and differ from those of threshold resummation by their $b=|\bf{b}|$ dependence 
\begin{eqnarray}
\label{eq:profile}
 P_{abd}(N,b,Q,\mu_F,\mu) = \nonumber &\\ 
 & \hspace{-2cm} \exp\left[E_a^{\rm PT}(N,b,Q,\mu_F,\mu)+E_b^{\rm PT}(N,b,Q,\mu_F,\mu) + F_d (N,Q,\mu)+ g_{abd} (N,\mu)\right]\,, 
\end{eqnarray}
with specific expressions given in appendix \ref{app:exponents}.
The parameter $\bar{\mu}$ in the third line of \eqref{eq:direct} acts as a cut-off on the recoil transverse momentum to avoid the singularity in the kinematic factor at ${\bf p}_T = {\bf Q}_T/2$, where the assumption that $Q_T$ is small compared to $p_T$ is not valid. This singularity is not present in the NLO calculation, but signals the case where the full transverse momentum of the photon in the CM frame is given by the recoil.
Note that the threshold-resummed result follows immediately from \eqref{eq:direct} if one neglects 
$\mathbf{Q}_T$ in the kinematic factor in the second line of
\eqref{eq:direct}, upon which the $\mathbf{Q}_T$ integral sets $\mathbf{b}$ to zero, and \eqref{eq:thresholddirect} is recovered. 

The joint-resummed expression for the fragmentation component is derived in analogy to the direct component~\cite{Laenen:2000ij,deFlorian:2005yj,Misra:2018nsy}. The result is similar to \eqref{eq:thresholdfrag}, with again $b$ dependence in the exponents, and with the kinematic factor linking recoil and threshold effects
\begin{eqnarray}
  \label{eq:frag}
  p_T^3   \frac{{\rm d} \sigma^{({\rm frag,joint})}_{AB\to \gamma + X} (x_T^2)}{{\rm d}p_T}
&=& \frac{p_T^4}{8 \pi S^2}\ \sum_{a,b,c}\  \int_{\cal C} \frac{{\rm d}N}{2 \pi i}\int \frac{{\rm d}^2 {\bf Q}_T }{(2\pi)^2}\;
\left( \frac{S}{4 |{\bf p}_T - {\bf Q}_T/2|^2}\right)^{N+1}\nonumber \\[1ex]
&&\hspace{3mm}  \times \hspace{2mm}  f_{a/A}(N,\mu_F) f_{b/B}(N,\mu_F)D_{\gamma/c}(2N+1,\mu_F) \nonumber \\[1ex]
&& \hspace{3mm} \times \int_0^1 {\rm d}\tilde{x}_T^2 (\tilde{x}_T^2)^{N} \frac{|\mathcal{M}_{ab\to c d}(\tilde{x}_T^2)|^2}{\sqrt{1-\tilde{x}_T^2}}C_{\delta}^{(ab\to c d)}(\alpha_s,\tilde{x}_T^2) \nonumber \\[1ex]
&\ & \hspace{3mm} \times \; 
\int {\rm d}^2 {\bf b}\; {\rm e}^{i {\bf b} \cdot {\bf Q}_T} \;
\theta\left(\bar{\mu}-|{\bf Q}_T|\right) P_{abcd}(N,b,Q,\mu_F,\mu)\,.
\end{eqnarray}
The avoidance of the singular case where $Q_T = 2p_T$ may in fact be treated alternatively to using a cut-off \cite{Sterman:2004yk}, with better numerical behaviour.

\subsection{Treatment of the kinematic singularity at $Q_T=2p_T$.}
\label{sec:treatment}
We shall avoid the singularity in the kinematic factor \eqref{eq:kinfactor} by the approximation proposed in \cite{Sterman:2004yk}
\begin{eqnarray}
\notag
\left(\frac{S}{4({\bf p}_T-{\bf Q}_T/2)^2}\right)^{N+1} & = & \left(\frac{4p_T^2}{S}\right)^{-N-1}\left(1-\frac{{\bf p}_T\cdot{\bf Q}_T}{p_T^2} + \frac{Q_T^2}{4p_T^2}\right)^{-N-1} \\[2ex]
\label{eq:expansion}
& \simeq & (x_T^2)^{-N-1}\exp\big[(N+1)\frac{{\bf p}_T\cdot{\bf Q}_T}{{p}_T^2}\left[1+\mathcal{O}\left(Q_T/p_T\right)\right]\big]\,.
\end{eqnarray}
Using  \eqref{eq:expansion} in   \eqref{eq:direct} and  \eqref{eq:frag}, one sees that the integral over ${\rm d}^2{\bf Q}_T$ produces the delta function $\delta\left({\bf b} - i(N+1){\bf p}_T/p_T^2\right)$. This delta function may be used to perform the integral over ${\rm d}^2\bf{b}$, which fixes $b=i(N+1)/p_T$.
\begin{figure}[t]
  \centering
  \includegraphics[width=0.6\textwidth]{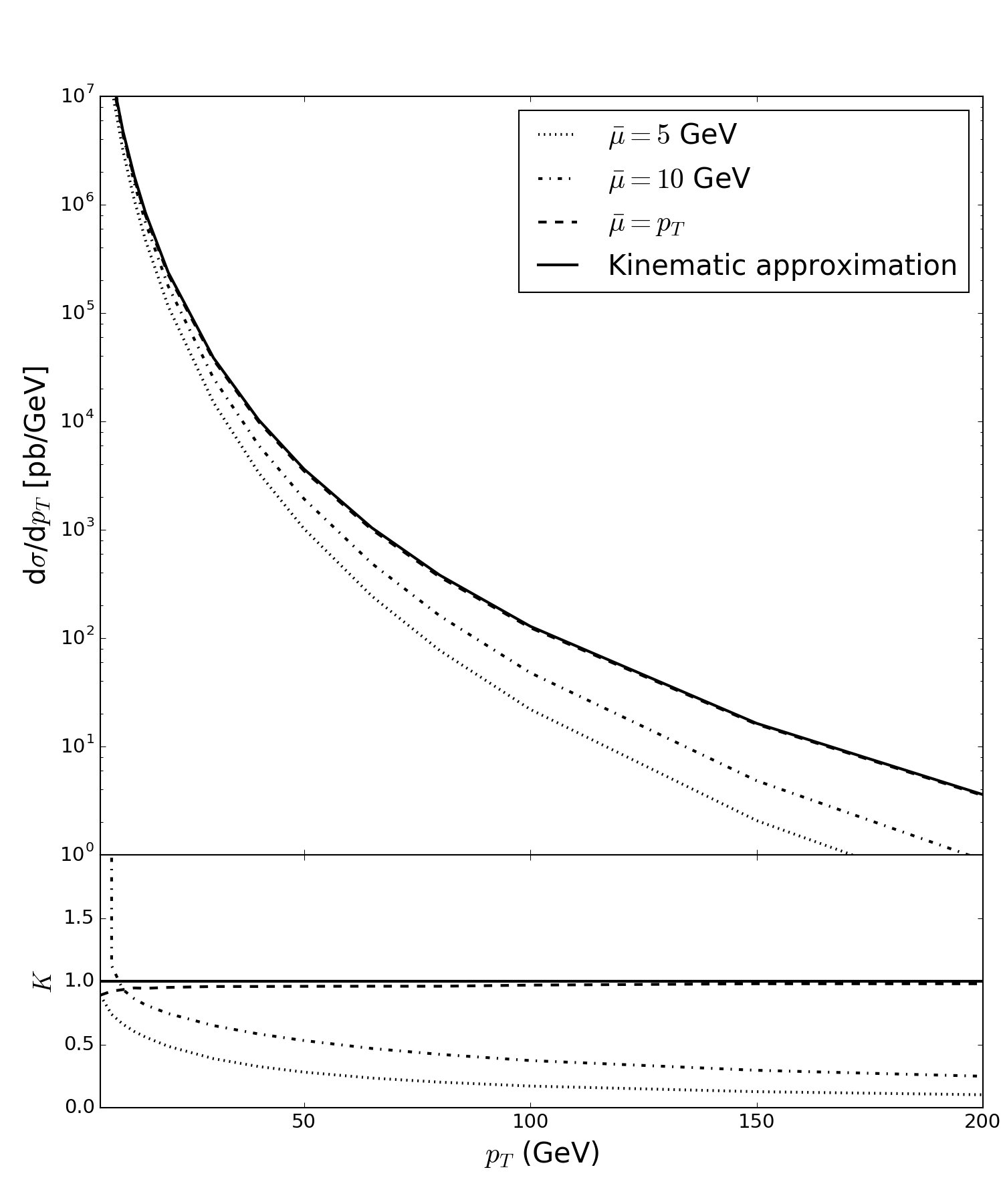}
  \caption{NLL joint-resummed $p_T$ distribution for different values of $\bar{\mu}$ compared to the approximation of the kinematic factor as given in \eqref{eq:expansion} (solid black line). The other lines correspond to $\bar{\mu} = 5$  GeV (dotted), $\bar{\mu}=10$  GeV (dash-dotted), and taking a variable cut-off of $\bar{\mu} = p_T$ (dashed). The ratio $K$ in the lower pane is of the cut-off results with respect to the kinematic approximation. The spike at very small $p_T$ in the $\bar{\mu}=10$ GeV curve is due to the singularity in the kinematic factor of \eqref{eq:frag}.}
  \label{fig:cutoff}
\end{figure}
The difference between this way of handling the kinematic singularity and using $\bar{\mu}$ can be seen in fig.~\ref{fig:cutoff}. The joint-resummed NLL cross-section for combined direct component and fragmentation component is shown for different values of $\bar{\mu}$, and for the approximated kinematic factor. One sees that at low fixed values of $\bar{\mu}$, there is a large  discrepancy between the two methods. The discrepancy decreases when using a higher value for $\bar{\mu}$, but one then runs into numerical instabilities for the calculation of the differential cross-section at small $p_T$. Alternatively one might use a $p_T$-dependent cut-off, such as $\bar{\mu}=p_T$. This  agrees much better with the approximation result. 
The approximation in \eqref{eq:expansion} has the significant added
benefit of being much more stable numerically, due to the smaller number of integrals, which are reduced from five for the cut-off method to one.
In our numerical results, presented in section \ref{sec:numerical-studies}, we therefore choose this method to handle the kinematic singularity. However, this choice does have some impact on the resummed expression, and leads to leading logarithms at NLP, as we now discuss.

The initial state exponent for joint resummation reads \cite{Laenen:2000de}
\begin{eqnarray}
\label{eq:initalsate}
E^{\rm PT}_a(N,b,Q,\mu_F,\mu) &=& \int_0^{Q^2}\frac{{\rm d}k_T^2}{k_T^2}A_a(\alpha_s(k_T^2))\left[J_0(bk_T)K_0\left(\frac{2Nk_T}{Q}\right)+\ln\left(\frac{\bar{N}k_T}{Q}\right)\right] \nonumber\\
\notag
&& - \ln\bar{N}\int_{\mu_F^2}^{Q^2} \frac{{\rm d} k_T^2}{k_T^2}A_a(\alpha_s(k_T^2)) \\
&\equiv& E^{\rm joint}_a(N,b,Q,\mu) + E^{\rm DGLAP}_a(N,Q,\mu_F,\mu),
\end{eqnarray}
where $\bar{N} = N {\rm e}^{\gamma_E}$ and $J_0$ and $K_0$ are Bessel functions. The function $A_a$ is written as a perturbative expansion in the strong coupling:
\begin{eqnarray}
A_a(\alpha_s) = \sum_{n=1}^{\infty}\left(\frac{\alpha_s}{\pi}\right)^n A^{(n)}_a\,,
\end{eqnarray}
with $A_a^{(1,2)}$ listed in appendix \ref{app:exponents}. The second line constitutes the large $N$ approximation of the DGLAP evolution. Note that the logarithm and the $K_0$ Bessel function in $E^{\rm joint}_a$ cancel in the limit $k_T\rightarrow 0$, since
\begin{eqnarray}
K_0(x) \stackrel{x\rightarrow 0}{=} -\ln\left(\frac{x{\rm e}^{\gamma_E}}{2}\right)\left[1+\mathcal{O}(x^2)\right]+\mathcal{O}(x^2).
\end{eqnarray} 
 Using that $J_0(ix) = I_0(x)$, the exponent $E_a^{\rm joint}$ with $b = i(N+1)/p_T$ reads
\begin{eqnarray}
E^{\rm joint}_{a}\left(N, b=i\frac{N+1}{p_T}, Q,\mu\right) & = & \int_0^{Q^2} \frac{{\rm d} k_T^2}{k_T^2} A_a(\alpha_s(k_T^2))\left[K_0\left(\frac{2Nk_T}{Q}\right) + \ln\left(\frac{\bar{N}k_T}{Q}\right) \right] \nonumber \\
&& \hspace{-2cm} + \int_0^{Q^2} \frac{{\rm d}k_T^2}{k_T^2}A_a(\alpha_s(k_T^2))\left[I_0\left(\frac{(N+1)k_T}{p_T}\right) 
- 1 \right]K_0\left(\frac{Nk_T}{p_T}\right)\,. \nonumber \\[2ex]
&\equiv& E^{\text{leading}}_a(N,Q,\mu) + E^{\rm recoil}_a(N,Q,\mu)\,.\label{eq:deform}
\end{eqnarray}
The term $E^{\text{leading}}_a$, when combined with $E^{\rm DGLAP}_a$ in   \eqref{eq:initalsate}, can be  recognized as the exponent for threshold resummation for the initial state. To see this, note that $E^{\rm leading}_a$ vanishes for small values of $x = \frac{2Nk_T}{Q}$; for larger values of $x$, the Bessel function $K_0(x)$ is exponentially suppressed. Since $K_0(x)$ is symmetric in $x$, we can write, to NLP accuracy,
\begin{eqnarray*}
K_0\left(\frac{2Nk_T}{Q}\right) + \ln\left(\frac{\bar{N}k_T}{Q}\right) \simeq \theta\left(\frac{\bar{N}k_T}{Q}-1\right)\ln\left(\frac{\bar{N}k_T}{Q}\right) + \mathcal{O}(1/\bar{N}^2)\,.
\end{eqnarray*}
The resulting expression for $E_a^{\text {thres}} = E^{\text {leading}}_a + E^{\rm DGLAP}_a$ is then
\begin{eqnarray}
\notag
E_a^{\rm thres}(N,Q,\mu_F,\mu) &=& -\int_{Q^2}^{Q^2/\bar{N}^2}\frac{{\rm d}k_T^2}{k_T^2}A_a\left(\alpha_s(k_T^2)\right)\ln\left(\frac{\bar{N}k_T}{Q}\right) \\
\label{eq:ISthres}
&& - \ln\bar{N}\int^{Q^2}_{\mu_F^2}\frac{{\rm d}k_T^2}{k_T^2}A_a\left(\alpha_s(k_T^2)\right). 
\end{eqnarray}
This can be rewritten so that the connection to threshold resummation is more transparent
\begin{eqnarray}
\notag
E_a^{\rm thres}(N,Q,\mu_F,\mu) &=& -\int^1_{1/\bar{N}}\frac{{\rm d}y}{y}\int^{y^2Q^2}_{\mu_F^2}\frac{{\rm d}k_T^2}{k_T^2}A_a\left(\alpha_s(k_T^2)\right) \\
\label{eq:cataniis}
&=& \int^1_{0}{\rm d}z \, \frac{z^{N-1}-1}{1-z}\int^{(1-z)^2Q^2}_{\mu_F^2}\frac{{\rm d}k_T^2}{k_T^2}A_a\left(\alpha_s(k_T^2)\right)\,,
\end{eqnarray}
where we used the NLL approximation \cite{Catani:2003zt}
\begin{eqnarray}
-\theta(1-z-1/\bar{N}) = z^{N-1}-1\,.
\end{eqnarray}
Carrying out the integrals yields
\begin{equation}
\label{eq:initialstate}
      E_a^{\rm thres}(N,Q, \mu_F,\mu)  =
\frac{1}{\alpha_s}h_a^{(0)} (\lambda) +
h_a^{(1)} (\lambda,Q,\mu_F,\mu)   \, ,
\end{equation}
where $\alpha_s \equiv \alpha_s(\mu^2)$, $\lambda \equiv \alpha_s b_0 \ln \bar{N}$ and the explicit forms of $h_a^{(0)}$ and $h_a^{(1)}$ are given in appendix \ref{app:exponents}.
We now turn our attention to the second term in \eqref{eq:deform}. 
As noted before, $b = i(N+1)/p_T$ is set to zero by neglecting ${\bf Q}_T$ in the kinematic factor, which sets $I_0\left(\frac{(N+1)k_T}{p_T}\right) = 1$. 
The term in the second line in \eqref{eq:deform} is therefore purely caused by a recoil, as it is only non-zero in the presence of ${\bf Q}_T$. To compute $E^{\rm recoil}_a$ we use leading order running of the strong coupling to change $\alpha_s(k_T^2)$ to $\alpha_s$ in~\eqref{eq:deform},
and expand the $I_0$ Bessel function (using $Q=2p_T$)  as 
\begin{eqnarray}
I_0\Big(x\,\big( 1+\frac{1}{N}\big)\Big) \overset{N\rightarrow\infty}{=} I_0\left(x\right) + \frac{x}{N}\,I_1\left(x\right)+ \mathcal{O}\left(\frac{1}{N^2}\right).
\end{eqnarray}
In \cite{Sterman:2004yk} only the first term in this expansion was used, here we need the second term as it gives rise to an NLP contribution. The leading asymptotic behaviour at large $N$ is given by
\begin{eqnarray}
E^{\rm recoil}_a(N, Q,\mu) &=& 2A^{(1)}_a\frac{\alpha_s}{\pi}\int_0^{2N}\frac{{\rm d}x}{x}\left(1+2\alpha_s b_0\ln\frac{x}{2N}\right)\Big[\left(I_0(x)-1\right)K_0(x)\nonumber\\
  && \hspace{1cm}  +\frac{x}{N}I_1(x)K_0(x)\Big]+\mathcal{O}\left(\frac{1}{N^2}\right)\nonumber \\
&\simeq& A^{(1)}_a\frac{\alpha_s}{2\pi}\left(\frac{\zeta(2)}{1-2\lambda}+\frac{\ln\bar{N}}{N}\right)
\equiv h^{(1)}_{a, {\rm recoil}}(\lambda,\alpha_s). \label{eq:recoilterm}
\end{eqnarray}
Thus, the approximation of the kinematic factor has led to a NNLL correction to the threshold initial state exponent at LP, but a LL correction at NLP. Note that these effects stem from (next-to-)soft \emph{non-collinear} momentum configurations of the emitted gluon and are only generated by a non-zero impact parameter $\bf{b}$. The contribution shows an NLP effect of non-collinear soft radiation in single-particle inclusive cross-sections.
The strictly collinear information is isolated in $E^{\rm thres}_a(N,Q,\mu_F,\mu)$. 
To summarize, the initial state exponent for joint resummation now reads
\begin{eqnarray}
    E^{\rm PT}_a(N,b=i\frac{N+1}{p_T},Q,\mu_F,\mu) &=& E^{\rm thres}_a(N,Q,\mu_F,\mu) +  E^{\rm recoil}_a(N,Q,\mu)\nonumber \\ &=&\frac{1}{\alpha_s}h^{(0)}_a(\lambda)+h^{(1)}_a(\lambda,Q,\mu_F,\mu)+h^{(1)}_{a,{\rm recoil}}(\lambda,\alpha_s).
    \label{eq:initialjoint}
\end{eqnarray}
We will use this expression for the NLL joint-resummed results presented in section \ref{sec:numerical-studies}. In the next section, we will see how we can modify $E^{\rm thres}_a(N,Q,\mu_F,\mu)$ to include NLP terms.

\subsection{Initial state threshold logarithms at next-to-leading power}
\label{sec:is}
To include initial state NLP terms in \eqref{eq:initialjoint} at LL accuracy we shall follow two approaches and subsequently compare them. In the first approach \cite{Kramer:1996iq} we modify the exponent by including $\mathcal{O}(1-z)$ corrections in the $z$-dependent integral of~\eqref{eq:cataniis}. In the second, following \cite{Kulesza:2002rh},  we rewrite the exponent in order to isolate a term that extends parton evolution to the complex scale $Q/\bar{N}$. Let us review these approaches.

For Drell-Yan and Higgs production it was argued in \cite{Basu:2007nu, Kramer:1996iq} that in order to include leading logarithmic corrections at NLP one may simply replace\footnote{In \cite{Basu:2007nu} an extra factor of $2$ was used in the NLP correction term for the case of a gluon emitter. As shown in \cite{DelDuca:2017twk}, this extra factor should not be there. }
\begin{eqnarray}
\label{eq:modificationIS}
\frac{z^{N-1}-1}{1-z}A^{(1)}_a \rightarrow 
\left(\frac{z^{N-1}-1}{1-z} - z^{N-1}\right)A^{(1)}_a
\end{eqnarray}
in the initial state exponent \eqref{eq:cataniis}, which then becomes
\begin{equation}
  \label{eq:9}
  E_a^{\rm thres} (N,Q,\mu_F,\mu) \equiv \frac{1}{\alpha_s}h^{(0)}_a(\lambda) + h^{(1)}_a(\lambda,Q,\mu_F,\mu) 
          +h'_a(\lambda,\alpha_s).
\end{equation}
Here, $h'_a$ is of order $1/N$, and is given in appendix \ref{app:exponents}. A recent LL resummation study at NLP for Drell-Yan \cite{Beneke:2018gvs} confirms this. As we discuss below, this choice reproduces at NLP the leading logarithms at NLO.

In the second approach the starting point instead is~\eqref{eq:ISthres}, which can be reorganized as
\begin{equation}
E^{\rm thres}_a(N,Q,\mu_F,\mu) = \int_{Q^2/\bar{N}^2}^{Q^2}\frac{{\rm d}k_T^2}{k_T^2}A_a(\alpha_s(k_T^2))\ln\left(\frac{k_T}{Q}\right) - \ln \bar{N}\int_{\mu_F^2}^{Q^2/\bar{N}^2}\frac{{\rm d}k_T^2}{k_T^2} A_a(\alpha_s(k_T^2))\,.\label{eq:extenevolution}
\end{equation}
One may add a term that is zero at NLL accuracy
\begin{eqnarray}
\label{eq:thresrewrite}
E^{\rm thres}_a(N,Q,\mu_F,\mu) &=& \int_{Q^2/\bar{N}^2}^{Q^2}\frac{{\rm d}k_T^2}{k_T^2}\left[A_a(\alpha_s(k_T^2))\ln\left(\frac{k_T}{Q}\right) - B_a(\alpha_s(k_T^2))\right] \nonumber \\
&&- \int_{\mu_F^2}^{Q^2/\bar{N}^2}\frac{{\rm d}k_T^2}{k_T^2} \left[A_a(\alpha_s(k_T^2))\ln \bar{N} + B_a(\alpha_s(k_T^2)\right] \nonumber \\
&\equiv& \hat{E}^{\rm thres}_a(N,Q,\mu) + E^{\rm evol}_a(N,Q,\mu_F,\mu)\,,
\end{eqnarray}
where the first order term of $B_{a}$ can be found in appendix \ref{app:exponents}. The first term is slightly different (indicated by the hat) from $E^{\rm thres}_a$ in \eqref{eq:initialstate}, which now reads
\begin{eqnarray}
\hat{E}^{\rm thres}_a(N,Q,\mu) = \frac{1}{\alpha_s}\hat{h}^{(0)}_a(\lambda) + \hat{h}^{(1)}_a(\lambda,Q,\mu)\,,
\end{eqnarray}
with $\hat{h}_a^{(0)}$ and $\hat{h}_a^{(1)}$ listed in appendix \ref{app:exponents}. The second term $E^{\rm evol}_a$ 
can be interpreted as extending evolution of the parton distribution functions from $\mu_F$ to the complex scale $Q/\bar{N}$,
since the large $N$ expansions of the diagonal splitting functions $P_{qq}(N)$ and $P_{gg}(N)$ are given by (see section \ref{sec:dif})
\begin{eqnarray}
P_{aa}(N) = -A_a(\alpha_s)\ln\bar{N}-B_a(\alpha_s) + {\cal O}(1/N)\,.
\end{eqnarray}
To now include NLP effects one can replace \cite{Kulesza:2002rh} the integrand of $E^{\rm evol}_a$ with the full splitting function and perform the extended evolution. In the next section we will verify that the LP NLL expression for $E^{\rm thres}_a(N,Q,\mu_F,\mu)$ will remain the same with this replacement. \\

\subsubsection{Similarity between the two approaches at LP}
\label{sec:dif}
To see the similarity between the two approaches, let us recall some technical aspects of evolution of the parton distribution functions (PDFs).  For our purposes we only need a brief description, 
more details can be found in \cite{Gluck:1989ze,Furmanski:1980cm, Furmanski:1982cw,Gluck:1993im}. 

The $N$-space DGLAP equations read, schematically,
\begin{eqnarray}
\frac{{\rm d}}{{\rm d}\ln \mu^2}E = \frac{\alpha_s(\mu^2)}{2\pi}P\,E\,,
\end{eqnarray}
where $E$ and $P$ are functions of $\alpha_s$ and $N$ in the case of the non-singlet (NS) evolution, and $2\times2$ matrices of such functions in the case of the singlet (S) evolution. This equation can be solved to each order in perturbation theory by writing $P(N)$ as a perturbative series
\begin{eqnarray}
P(N) = P^{(0)}(N) + \frac{\alpha_s(\mu^2)}{2\pi}P^{(1)}(N) + {\cal O}(\alpha_s^2)\,.
\end{eqnarray}
with the explicit $N$-space expressions for the one and two-loop splitting functions $P^{(0)}(N)$ and $P^{(1)}(N)$ given in appendix B of \cite{Floratos:1981hs}.
PDF evolution is encoded in the evolution matrices $E$
\begin{eqnarray}
q_{NS}(N, \mu^2) &=& E_{NS}(\mu^2, \mu_0^2)\, q_{NS}(N, \mu_0^2)\,, \nonumber \\
q_{S}(N, \mu^2) &=& E_{qq}(\mu^2, \mu_0^2)\,q_{S}(N, \mu_0^2) + E_{qg}(\mu^2, \mu_0^2)\,g(N, \mu_0^2)\,, \nonumber \\
g(N, \mu^2) &=& E_{gq}(\mu^2, \mu_0^2)\,q_{S}(N, \mu_0^2) + E_{gg}(\mu^2, \mu_0^2)\, g(N, \mu_0^2)\,, 
\end{eqnarray}
where $\mu_0$ and $\mu$ are the initial and final scale of evolution, respectively.
To compare the modified exponent (ME) and the PDF evolution approach, let us first examine the non-singlet case. The extended NS evolution at NLO accuracy of an initial state parton from the factorization scale $\mu_F$ to the complex scale $Q/\bar{N}$ reads
\begin{eqnarray}
\nonumber
E_{\rm NS}(Q^2/\bar{N}^2, \mu_F^2) &\stackrel{{\rm NLO}}{=}& \exp\left[-\frac{\alpha_s(Q^2/\bar{N}^2) - \alpha_s(\mu_F^2)}{4 \pi^2 b_0}\left\{P^{(1)}_{\rm NS}(N) - \frac{2 \pi b_1}{b_0} P^{(0)}_{\rm NS}(N) \right\}\right]\\
&& \times \exp\left[-\frac{P^{(0)}_{\rm NS}(N)}{2\pi b_0}\ln\left(\frac{\alpha_s(Q^2/\bar{N}^2)}{\alpha_s(\mu_F^2)}\right)\right]. \label{eq:nlons}
\end{eqnarray}
The running of $\alpha_s$ is governed by the beta function
\begin{eqnarray}
\mu^2 \frac{{\rm d}\alpha_s(\mu^2)}{{\rm d}\mu^2} = \beta(\alpha_s(\mu^2)) = -\alpha_s^2 \sum_{k=0}^{\infty}\alpha_s^k b_k.
\label{eq:betaeqn}
\end{eqnarray}
The explicit forms of the first two coefficients $b_0$ and $b_1$ are given in appendix \ref{app:exponents}. To NLL accuracy we must use the NLO beta function (i.e. including $b_0$ and $b_1$) for the LO evolution exponent (the second line of \eqref{eq:nlons}). The LO beta function (including $b_0$ only) is sufficient to NLL accuracy for the first line of \eqref{eq:nlons}. This results in
\begin{eqnarray}
\alpha_s(Q^2/\bar{N}^2)-\alpha(\mu_F^2) &\stackrel{\text{LO beta function}}{=}& \frac{\alpha_s}{1+\alpha_s b_0 \ln\left(Q^2/(\bar{N}\mu)^2\right)} - \frac{\alpha_s}{1+\alpha_s b_0 \ln\left(\mu_F^2/\mu^2\right)} \nonumber  \\
&=& \frac{\alpha_s}{1-2\lambda} - \alpha_s - \alpha_s^2 b_0 \ln\left(\frac{Q^2}{\mu_F^2}\right)+\mathcal{O}(\alpha_s^3).
\label{eq:asLO}
\end{eqnarray}
One observes that the LO running of the strong coupling introduces a dependence on the scale that is of NNLL accuracy and only vanishes for  
$Q=\mu_F$. We therefore truncate \eqref{eq:asLO} to NLL accuracy
\begin{eqnarray}
\alpha_s(Q^2/\bar{N}^2)-\alpha(\mu_F^2) &=& \frac{\alpha_s}{1-2\lambda} - \alpha_s,
\end{eqnarray}
resulting in the following expression for the NS evolution matrix
\begin{eqnarray}
\nonumber
E_{\rm NS}(Q^2/\bar{N}^2, \mu_F^2) 
&\stackrel{{\rm NLL}}{=}& \exp\left[\frac{\alpha_s}{2 \pi}\frac{2 \lambda }{1-2\lambda}\left\{\frac{b1}{b_0^2}P^{(0)}_{\rm NS}(N)- \frac{1}{2 \pi b_0}P^{(1)}_{\rm NS}(N)\right\}\right] \\
&& \times \exp\Bigg[\frac{P^{(0)}_{\rm NS}(N)}{2\pi b_0} \ln(1-2\lambda) + \frac{P^{(0)}_{\rm NS}(N)}{2\pi b_0}\alpha_s
\nonumber \\
&& \times \left\{\frac{b_1}{b_0}\frac{\ln(1-2\lambda)}{1-2\lambda} + \frac{2\lambda}{1-2\lambda}b_0\ln\left(\frac{Q^2}{\mu^2}\right)+b_0\ln\left(\frac{Q^2}{\mu_F^2}\right)\right\}\Bigg].\label{eq:totalres}
\end{eqnarray}
We must also expand the functions $P^{(0)}_{\rm NS}(N)$ and $P^{(1)}_{\rm NS}(N)$ (given in appendix B of \cite{Floratos:1981hs}) to NLL accuracy. The large $N$ approximation of $P^{(0)}_{\rm NS}$ is given by
\begin{eqnarray}
\label{eq:pns0}
P^{(0)}_{{\rm NS}}(N) \stackrel{N\rightarrow \infty}{=} - 2A^{(1)}_q\ln\bar{N} - 2B^{(1)}_q +\mathcal{O}\left(\frac{1}{N}\right).
\end{eqnarray} 
To get an expression for $E_{\rm NS}(Q^2/\bar{N}^2, \mu_F^2)$ accurate to NLL order, the $P^{(0)}_{\rm NS}$ factor in the second exponent of \eqref{eq:totalres}, i.e. the one that is not multiplied with $\alpha_s$, will be expanded to $\mathcal{O}(1)$. The other $P^{(0)}_{\rm NS}$ factor will be expanded to $\mathcal{O}(\ln\bar{N})$. For $P^{(1)}_{\rm NS}(N)$ we only need the $\mathcal{O}(\ln\bar{N})$ term 
\begin{eqnarray}
P^{(1)}_{\rm NS}(N) \stackrel{N\rightarrow \infty}{=} -4A^{(2)}_q\ln\bar{N}.
\end{eqnarray}

Now we can use these expansions in   \eqref{eq:totalres}, which gives
\begin{eqnarray}
\nonumber
\ln\left(E_{\rm NS}(Q^2/\bar{N}^2, \mu_F^2) \right) &\stackrel{{\rm NLL}}{=}& -\frac{A^{(1)}_q}{\pi b_0}\ln\bar{N}\ln\left(1-2\lambda\right) - \frac{B^{(1)}_q}{\pi b_0}\ln\left(1-2\lambda\right)  \\
\nonumber
&& +\alpha_s\Big[- \frac{A^{(1)}_q b_1}{b_0^2\pi}\frac{\ln\left(1-2\lambda\right)}{1-2\lambda}\ln\bar{N}  - \frac{A^{(1)}_q b_1}{\pi b_0^2}\frac{2\lambda}{1-2\lambda}\ln\bar{N} \\
&& \hspace{-25mm} - \frac{A^{(1)}_q}{\pi}\frac{2\lambda}{1-2\lambda}\ln\frac{Q^2}{\mu^2}\ln\bar{N} 
\label{eq:difference}
 - \frac{A^{(1)}_q}{\pi} \ln\bar{N} \ln\frac{Q^2}{\mu_F^2} + \frac{A^{(2)}_q}{b_0 \pi^2}\frac{2\lambda}{1-2\lambda}\ln\bar{N}\Big] .
\end{eqnarray}
Adding this result to $\hat{E}^{\rm thres}_a(N,Q,\mu)$ in \eqref{eq:thresrewrite} results in the correct expression for $E^{\rm thres}_a(N,Q,\mu_F,\mu)$ at NLL accuracy. 
For singlet evolution we can arrive at a similar conclusion when one neglects the off-diagonal contributions. Therefore, we find that there are no spurious terms created at LP NLL accuracy by replacing $E^{\rm evol}_a(N,Q,\mu_F,\mu)$ in \eqref{eq:thresrewrite} with an extended evolution of the PDFs.

\subsubsection{Differences between methods at the next-to-leading power level}

Having shown the equivalence of both methods to NLL accuracy at LP we can now include NLP terms in each, and compare them. For the ME approach, we include the NLP terms via \eqref{eq:modificationIS}. We will refer to this as option 1. We have several options in the extended PDF evolution approach. Let us start with the non-singlet case. The first option (2a) is to upgrade the large $N$  splitting functions by adding the $\mathcal{O}(1/N)$ term in $P^{(0)}_{\rm NS}$ 
\begin{eqnarray*}
\exp\left[\frac{P_{\rm NS}^{(0)}(N)}{2\pi b_0}\ln(1-2\lambda)\right] = \exp\left[\frac{1}{2\pi b_0}\left(-2A_q^{(1)}\ln\bar{N}-2B_q^{(1)}-\frac{A_q^{(1)}}{N}\right)\ln(1-2\lambda)\right].
\end{eqnarray*}
A brief calculation shows that this results in exactly the same $\ln N/N$ term as for the ME approach, which is encoded in 
$h_a'$ of \eqref{eq:9}.
In addition (option 2b) we can upgrade $P^{(1)}_{\rm NS}$ with its $\mathcal{O}(1/N)$ approximation. Finally (option 2c), we can keep the full form of $P^{(0)}_{\rm NS}$ and $P^{(1)}_{\rm NS}$. Instead of truncating the evolution of the strong coupling to NLL order, we will also use the NLO evolution of $\alpha_s$ in option 2b and 2c. \\

In the singlet case, the off-diagonal components of the splitting matrix vanish  if one expands the splitting matrices $P^{(0,1)}_S$ and ignores the terms of $\mathcal{O}(1/N)$. This leads to the same result as in the ME approach. A complication arises if we want to expand the splitting matrices to $\mathcal{O}(1/N)$; one must solve a coupled set of equations, leading to an equation for the evolution matrix that is not exponentiated. This can already be seen at the LO level, where the solution is of the form
\begin{eqnarray}
E_{\rm S}^{(0)}(Q^2/\bar{N}^2,\mu_F^2) &=& \exp\left[-\frac{P^{(0)}_{\rm S}(N)}{2\pi b_0}\ln\left(\frac{\alpha_s(Q^2/\bar{N}^2)}{\alpha_s(\mu_F^2)}\right)\right]\nonumber  \\
&& \hspace{-1.5cm}= e_1\, \exp\left[-\frac{\lambda_1}{2\pi b_0}\ln\left(\frac{\alpha_s(Q^2/\bar{N}^2)}{\alpha_s(\mu_F^2)}\right)\right] + e_2\, \exp\left[-\frac{\lambda_2}{2\pi b_0}\ln\left(\frac{\alpha_s(Q^2/\bar{N}^2)}{\alpha_s(\mu_F^2)}\right)\right].
\end{eqnarray}
Here, $\lambda_{1,2}$ are the eigenvalues of $P^{(0)}_S$ and $e_{1,2}$ are projection matrices defined as
\begin{eqnarray}
e_{1,2} = \frac{1}{\lambda_{1,2}-\lambda_{2,1}}\left[P_S^{(0)}-\lambda_{2,1}I\right].
\end{eqnarray}
By including the off-diagonal components of the splitting functions, one allows for the emission of (soft) quarks. This changes the identity of the emitter and is not in the spirit of soft gluon resummation, but it does give rise to NLP contributions \cite{vanBeekveld:2019prq, Moult:2019mog}. Finally, also for the singlet case we have the option to evolve with the full splitting matrices. \\

To summarize, in section \ref{sec:numerical-studies} we shall study four approaches to include NLP terms:
\begin{enumerate}
\item[1)] A modified initial state resummation coefficient, where  $h'_a(\lambda,\alpha_s)$ is added to \eqref{eq:initialstate} in the case of threshold resummation, and \eqref{eq:initialjoint} in the case of joint resummation. 
\item[2a)] Perform an extended evolution of the initial state partons from $\mu_F$ to $Q/\bar{N}$, keeping only the diagonal terms in the splitting functions. The full exponential form of the evolution equation is kept (like in \eqref{eq:totalres}). The coupling constant is evolved with the NLO beta function for the term proportional to $\ln\left( \frac{\alpha_s(Q^2/\bar{N}^2)}{\alpha_s(\mu_F^2)}\right)$ and with the LO beta function for the term proportional to $\alpha_s(Q^2/\bar{N}^2) -\alpha_s(\mu_F^2)$. Similarly, $P^{(0)}$ is expanded to $\mathcal{O}(1/N)$ and $P^{(1)}$ is expanded to $\mathcal{O}(\ln \bar{N})$. Note that this method should give the same result as the ME approach only for $\mu_F = Q$, as the LO beta function introduces a dependence on the scales which is of NNLL order (see \eqref{eq:asLO}). 
\item[2b)] Evolve the partons from $\mu_F$ to $Q/\bar{N}$, including the non-diagonal terms in the singlet case. Expand the splitting functions to $\mathcal{O}(1/N)$. The NLO beta function is used to evolve $\alpha_s$. 
\item[2c)] Evolve the partons from $\mu_F$ to $Q/\bar{N}$ and keep the full form of all splitting functions. The NLO beta function is used to evolve $\alpha_s$. 
\end{enumerate}
Let us comment on what NLP terms we take into account with each option. In option 1 and 2a we include the effect of next-to-soft collinear gluons that are emitted from either one of the initial state partons, as we only include the $\mathcal{O}(1-z)$ part of the diagonal splitting function. With option 2b (and 2c) we also allow for the emission of collinear quarks. This results in an NLP LL contribution when the quark momentum becomes soft-collinear \cite{vanBeekveld:2019prq, Moult:2019mog}. As the full NLO beta function (i.e. including $b_0$ and $b_1$ in \eqref{eq:betaeqn}) is used to evolve $\alpha_s$, we also include terms that are beyond NLP LL order in option 2b (and 2c). All these options only include the effects of collinear parton emission. For our joint resummation results we also include an NLP LL term \eqref{eq:recoilterm}, created by non-collinear gluon emissions, in each of the options.  

\subsection{Final state next-to-leading power terms}
\label{sec:fs}
For the non-fragmenting final state parton, we include NLP terms by a modification of the final state exponent, as shown in \eqref{eq:modificationIS}. The final state exponent then reads \cite{Mathews:2004pu}
\begin{equation}
  \label{eq:101}
  F_d (N,\mu,Q) \equiv \frac{1}{\alpha_s}f^{(0)}_d(\lambda) + f^{(1)}_d(\lambda,Q,\mu) 
          +f'_d(\lambda,\alpha_s),
\end{equation}
with each of the functions $f_d^{(0)}, f_d^{(1)}$ and $f'_d$ listed in appendix \ref{app:exponents}.
For the fragmenting final state parton we will compare again two ways of taking into account the NLP terms. \\
In option 1 we modify the final state exponent for parton~$c$, in analogy to parton~$d$ \cite{Basu:2012ma}. This results in an exponent for the fragmenting final state parton of the form
\begin{equation}
\label{eq:46}
  E_c^{\mathrm{PT}}(N,Q,\mu_F, \mu) = \frac{1}{\alpha_s} h^{(0)}_c(\lambda) + h^{(1)}_c(\lambda,Q,\mu_F, \mu) + h'_c(\lambda,\alpha_s)\,.
\end{equation}
The functions $h^{(0)}_c$, $h^{(1)}_c$ and $h'_c$ are the same ones as in \eqref{eq:9}. This approach will contain the leading NLP terms stemming from gluons radiated from the fragmenting leg. 

In our second approach (2a) we include the NLP effects through an extended evolution of the fragmentation function. The additional complication here  is that the evolution equation for the photon fragmentation function is not homogeneous. To leading order the non-singlet evolution equation for $D_{\gamma/c}(N,\mu_F)$ is given by
\begin{eqnarray}
\label{eq:diffinhomo}
\frac{{\rm d}D_{\gamma/c}(N,\mu^2)}{{\rm d}\mu^2} = k(N) + P(N)\,D_{\gamma/c}(N,\mu^2),
\end{eqnarray}
where the Mellin-space photon-parton splitting functions $k(N)$ and the purely partonic splitting functions $P(N)$ are 
\begin{eqnarray}
k(N) = \frac{\alpha}{2\pi}k^{(0)}(N)+\frac{\alpha\alpha_s(Q^2)}{(2\pi)^2}k^{(1)}(N)\,, \\
P(N) = \frac{\alpha_s(Q^2)}{2\pi}P^{(0)}(N)+\left(\frac{\alpha_s(Q^2)}{2\pi}\right)^2P^{(1)}(N)\,,
\end{eqnarray}
where $\alpha$ is the electromagnetic fine structure coupling. The explicit forms of $k^{(i)}(N)$ are given in \cite{PhysRevD.45.3986}. To solve the  differential equation in \eqref{eq:diffinhomo} one can split
$D_{\gamma/c}(N,\mu^2)$ into a homogeneous part and an inhomogeneneous point-like term $D^{{\rm PL}}_{\gamma/c}(N,\mu^2)$.
The inhomogeneous term will result in a correction that is of NLP order. To see this consider the NLO NS solution of the point-like (inhomogeneous) term \cite{PhysRevD.45.3986}
\begin{eqnarray}
D^{{\rm PL}}_{\gamma/c}(N,\mu^2) &=& \frac{4\pi}{\alpha_s(\mu^2)}\left\{1+\frac{\alpha_s(\mu^2)}{2\pi}U\right\}\left[1-L^{1-\frac{P^{(0)}}{2\pi b_0}}\right]\frac{1}{1-\frac{P^{(0)}}{2\pi b_0}}\frac{\alpha}{8\pi^2b_0} k^{(0)} \\
&& + \left[1-L^{-\frac{P^{(0)}}{2\pi b_0}}\right]\frac{1}{-P^{(0)}}\frac{\alpha}{2\pi}\left[k^{(1)}-\frac{2\pi b_1}{ b_0}k^{(0)}-Uk^{(0)}\right] + \mathcal{O}(\alpha_s),
\end{eqnarray}
where $L^{\frac{P^{(0)}}{2\pi b_0}} \equiv \left(\frac{\alpha_s(\mu^2)}{\alpha_s(\mu_0^2)}\right)^{\frac{P^{(0)}}{2\pi b_0}}$, and $P^{(0)}$, $U$, $k^{(0)}$ and $k^{(1)}$ are all functions of the Mellin moment $N$. For the NS case, $P^{(0)}$ and $U$ commute and $U$ is given by
\begin{eqnarray}
U = \frac{b_1}{b_0^2}P^{(0)} - \frac{1}{2\pi b_0}P^{(1)}.
\end{eqnarray}
Defining $\frac{P^{(0)}}{2\pi b_0} \equiv \mathcal{P}$ and writing as usual $\alpha_s \equiv \alpha_s(\mu^2)$ , the inhomogeneous term can be written as
\begin{eqnarray}
D^{{\rm PL}}_{\gamma/c}(N,\mu^2) &=& \frac{\alpha k^{(0)}}{2\pi b_0}\Bigg\{\frac{1-L^{1-\mathcal{P}}}{1-\mathcal{P}}\left(\frac{1}{\alpha_s}+\frac{1}{2\pi}\left(\frac{2\pi b_1}{b_0}\mathcal{P}-\frac{P^{(1)}}{2\pi b_0}\right)\right)\\
&& + \frac{1}{2\pi}\frac{1-L^{-\mathcal{P}}}{\mathcal{P}}\left((1+\mathcal{P})\frac{2\pi b_1}{b_0}-\frac{k^{(1)}}{k^{(0)}}-\frac{P^{(1)}}{2\pi b_0}\right)\Bigg\}.
\end{eqnarray}
The factor $k^{(0)}$ is proportional to $\frac{1}{N}$, while $k^{(1)}$ starts at order $1/N$ for the quark case and at $\mathcal{O}(1/N^2)$ in the gluon case. Substituting $\mu_0 = \mu_F$ and $\mu = Q/\bar{N}$ and expanding this equation for $\alpha_s b_0 \ln\bar{N} \rightarrow 0$ gives
\begin{eqnarray}
\label{eq:plevol}
D^{{\rm PL}}_{\gamma/c}(N,Q^2) &\simeq& -\frac{\alpha k^{(0)}}{\pi}\ln{\bar{N}} = -\frac{\alpha}{\pi N}\ln{\bar{N}}.
\end{eqnarray}
This shows that the extended evolution will give us terms that are of the same order as the leading NLP terms in the homogeneous part. By adding \eqref{eq:plevol} to the homogeneous solution, one can include the $\ln N /N$ effects of the fragmenting final state. The homogeneous term will be treated as in option 2a for the initial state. 
The second option in this approach is to evolve the fragmentation function to the complex scale $Q/\bar{N}$ with the full evolution equation. We call this option 2c, as it is closely related to option 2c to include the NLP terms for the initial state. To summarize, we will compare the following ways of including the NLP terms for the fragmenting final state:
\begin{enumerate}
\item[1)] Modification of the final state exponent \eqref{eq:46}.
\item[2a)] Using \eqref{eq:plevol} for the inhomogeneous term and using option 2a of the initial state for the homogeneous term.
\item[2c)] Performing the full evolution of the fragmentation functions from $\mu_F$ to $Q/\bar{N}$.
\end{enumerate}

In the last two sub-sections we have indicated different approaches of including NLP terms for the joint or threshold resummed production of prompt photons. A listing of which LL NLP results are captured at NLO by our resummed expression is given in appendix \ref{app:NLOcomparison}. The LL NLP terms that we do not catch can be classified in three categories. First, some of the terms have a non-collinear origin \cite{vanBeekveld:2019prq}. These cannot be reproduced via a modification of the splitting/fragmentation functions, as these only contain collinear effects. A second category follows from the modification of the hard scattering kinematics as a result of the next-to-soft gluon emission \cite{DelDuca:2017twk,vanBeekveld:2019prq}. The $\mathcal{O}(1-z)$ expansion of the Born function is then multiplied with a LL LP term, which results in an NLP logarithm. There terms are not included in the present study. A third set of LL NLP terms not included is due to  NLP phase space effects. However, we do capture all LL NLP effects that have a collinear origin for both gluon and quark emission.   \\
In the next section we will numerically examine the effects of including the (next-to-soft) collinear NLP terms via the different approaches. Note that whenever we will use option 2b to include the initial state NLP terms, we will use option 2a to include the final state NLP terms unless explicitly stated otherwise.

\section{Numerical studies}
\label{sec:numerical-studies}

In this section we perform the numerical assessment of the various ways of including NLP terms we discussed in the previous section. For this we consider the case of the LHC operating at a center-of-mass energy of $\sqrt{S} = 13$ TeV. For the parton distributions we use the central MMHT set \cite{Harland-Lang:2014zoa}, corresponding to $\alpha_s(M_Z^2) = 0.120$. For the fragmentation function, which are rather poorly known \cite{Kaufmann:2017lsd,Catani:2018krb}, we use the results of \cite{Gluck:1993zx}. We evolve the PDFs, the fragmentation function and the strong coupling in $N$-space using the code of \cite{Vogt:2004ns} in the variable flavor number scheme. Further choices for the input parameters are as follows: $m_c =1.4\,\mathrm{GeV}$, $m_b =4.5 \, \mathrm{GeV}$, $m_t = 175 \, \mathrm{GeV}$ and we set the factorization scale $\mu_F$ equal to the renormalization scale $\mu$. We checked that the results presented below do not change significantly when $\mu_F \neq \mu$. The hard scale $Q$ is set to $2p_T$ and $\mu$ is chosen equal to $Q$ unless stated otherwise. For the LP direct threshold-resummed $p_T$ distributions we use \eqref{eq:thresholddirect}, and for the LP fragmentation results we use \eqref{eq:thresholdfrag}. As shown in section \ref{sec:treatment}, the approximation of the kinematic factor in \eqref{eq:expansion} results in a recoil term that is added to the initial state threshold exponent for the joint-resummed results (see \eqref{eq:initialjoint}). The various ways to include initial and final state NLP effects are described in sections \ref{sec:is} and \ref{sec:fs}.  \\

\begin{figure}
  \centering
  \includegraphics[trim=0 0 0 0,clip,width=0.55\textwidth]{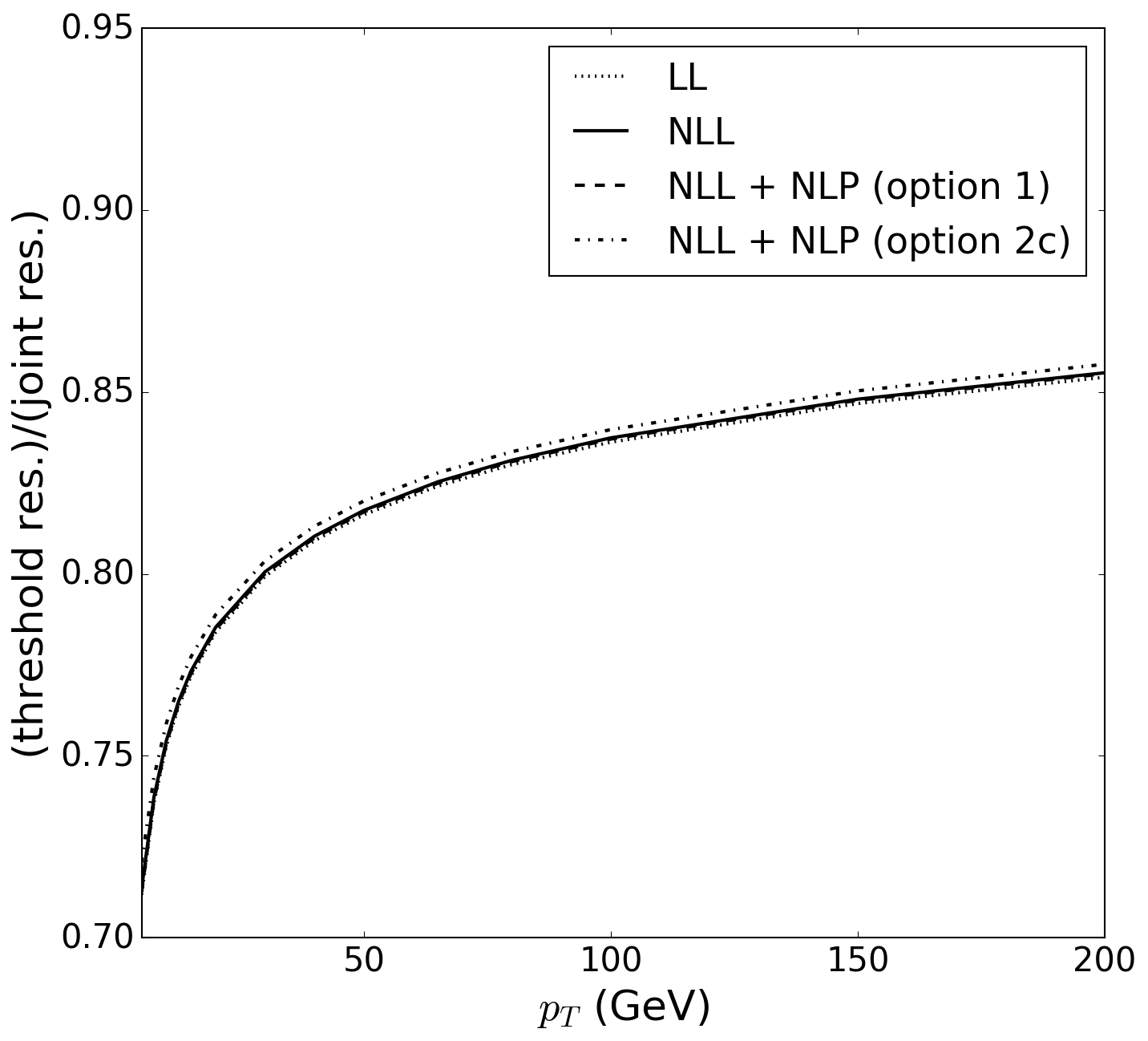}
  \caption{Ratio of the threshold-resummed to the joint-resummed combined $p_T$ distribution. Four different levels of accuracy are shown: LP LL (dotted line), LP NLL (solid line), LP NLL + NLP included as option 1 described in section \ref{sec:is} and \ref{sec:fs} (dashed line) and LP NLL + NLP included as option 2c (dash-dotted line).}
  \label{fig:JRoTR}
\end{figure}

\begin{figure}
  \centering
  \includegraphics[trim=0 0 0 0,clip,width=\textwidth]{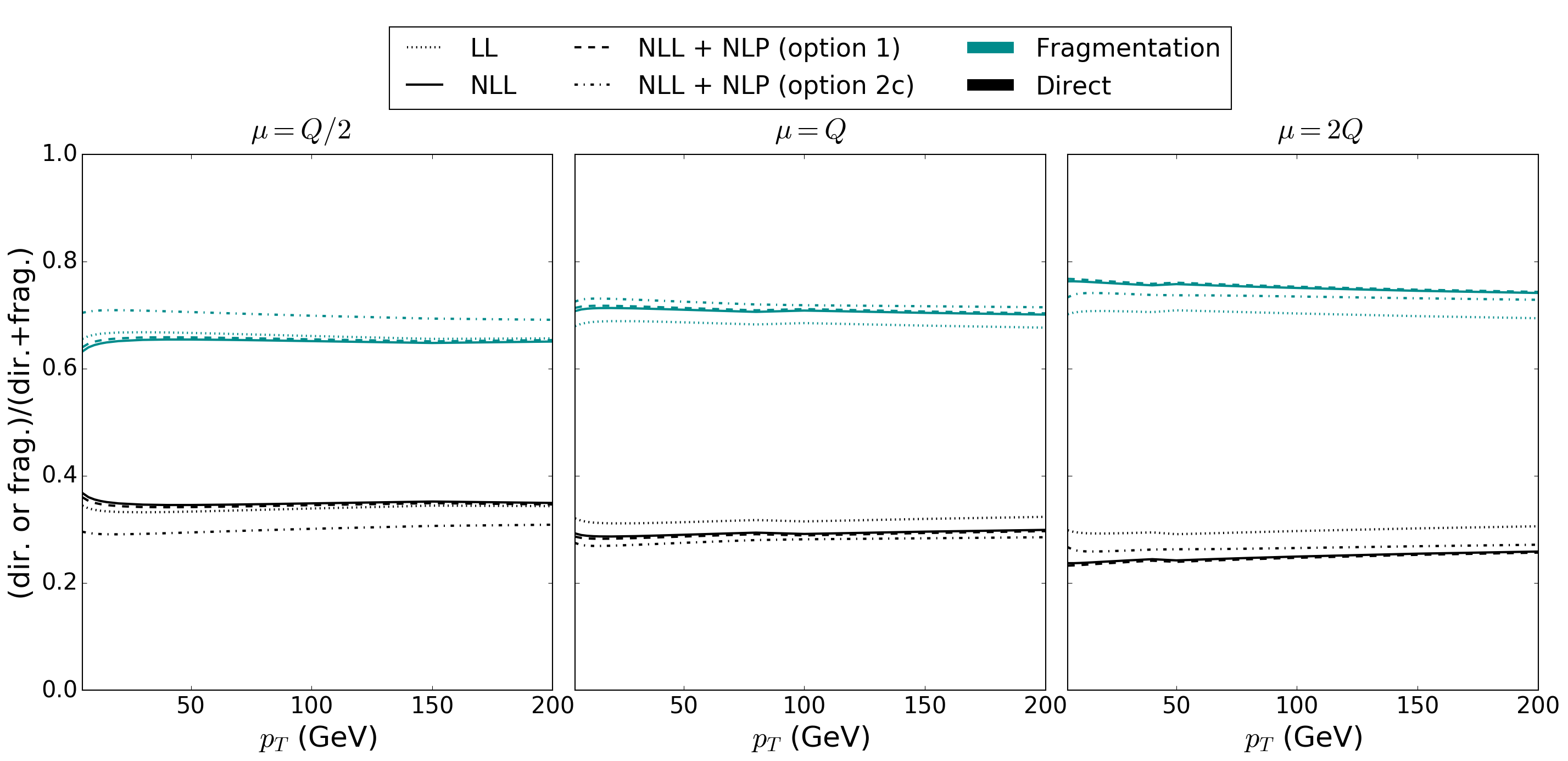}
  \caption{Relative contribution of the direct (black) and fragmentation (blue) component to the combined joint-resummed $p_T$ distribution for $\mu = Q/2$ (left), $\mu = Q$ (middle) and $\mu = 2 Q$ (right). }
  \label{fig:dirvsfrag}
\end{figure}

We begin by showing in fig.~\ref{fig:JRoTR} the ratio of the threshold-resummed to the joint-resummed combined $p_T$ distribution for four levels of accuracy. One observes that the ratio of threshold to joint resummation lies between 0.7 and 0.9, i.e. the $p_T$ distribution is enhanced by joint resummation by about $ 10-30\%$ with respect to threshold resummation. The difference between the two distributions decreases for higher $p_T$ values. The ratio does not change significantly under variations of $\mu$. \\
In fig.~\ref{fig:dirvsfrag} we show the relative contributions of the direct and fragmentation components to the combined joint-resummed $p_T$ distribution for three scale choices and different levels of accuracy. The fragmentation component dominates over the direct component for all $p_T$ values and scales. For the LP NLL and the NLL + NLP (included via option 1) cases, the relative contribution of the fragmentation component increases for higher values of $\mu$. On the other hand, the LP LL and LP NLL + NLP (option 2c) ratios are barely affected by the choice of scale. 
\subsection{Initial state NLP terms}
\begin{figure}
  \centering
  \includegraphics[trim=0 0 0 0,clip,width=\textwidth]{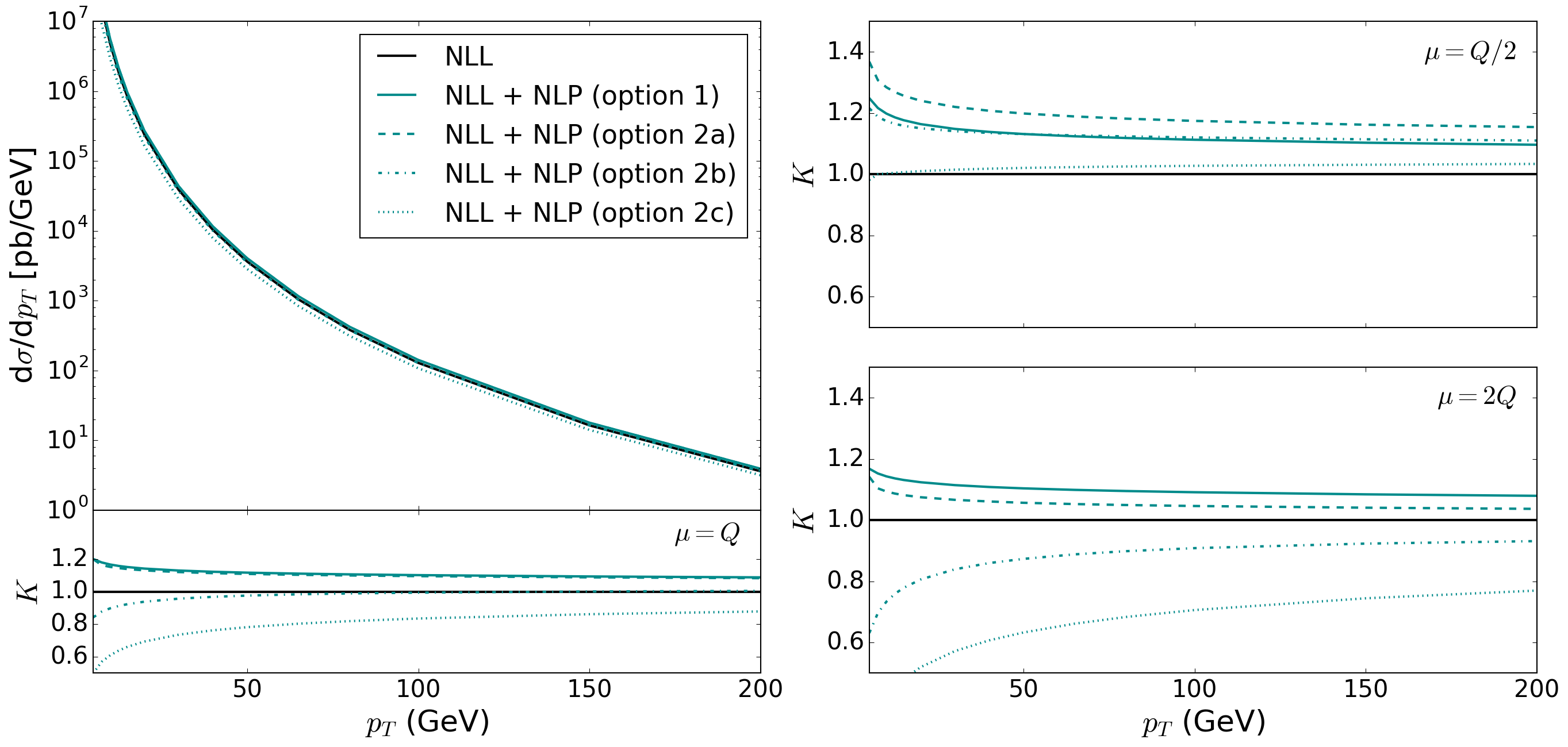}
  \caption{Combined joint-resummed $p_T$ distribution for LP resummation (black) and the inclusion of NLP effects for the initial state (blue) for different levels of accuracy as listed in section \ref{sec:is}. The bottom pannel and the plots on the right show the ratio $K$ with respect to the LP NLL resummed result for three choices of scale $\mu =  Q$ (left), $\mu =  Q/2$ (top right) and $\mu =  2 Q$ (lower right).}
  \label{fig:ISNLPJR}
\end{figure}

We show in fig.~\ref{fig:ISNLPJR} the results for the combined joint-resummed $p_T$ distribution, where initial state NLP terms are included with the various options listed in section~\ref{sec:is}. The final state NLP terms are not included here. One observes that the ME approach (option 1) and the diagonal extended evolution of the PDFs (option 2a) give the same correction at $\mu = Q$, as explained in item 2a in section~\ref{sec:is}. The ratio of option 1 to option 2a shows a slight dependence on the scale, which is caused by the NNLL scale dependent terms in \eqref{eq:asLO}. The NLP terms included via either option 1 or option 2a correspond to a correction to the combined $p_T$ distribution of about  $10\%$ at large $p_T$ (close to threshold), and about $20\%$ at lower $p_T$ values. These numbers hold for both the direct and fragmentation components at the central scale.  \\
The off-diagonal extended evolution of the PDFs to NLP LL accuracy (option 2b) diminishes the NLP contribution for a central scale of $\mu = Q$. This is caused by a relative sign difference between the diagonal and off-diagonal $\mathcal{O}(1/N)$ terms of the singlet splitting functions. This relative contribution is however strongly dependent on the scale. For a scale choice of $\mu = Q/2$, one observes a positive contribution from the extended evolution, while for $\mu = 2 Q$ the relative contribution is negative. When we allow for the full form of the splitting functions (option 2c), we see a large correction of $-40\%$ for small $p_T$ values and $-10\%$ for larger $p_T$ for $\mu = Q$. Again this relative correction has a large scale dependence. 
\subsection{Final state NLP terms}
\begin{figure}
  \centering
  \includegraphics[trim=0 0 0 0,clip,width=\textwidth]{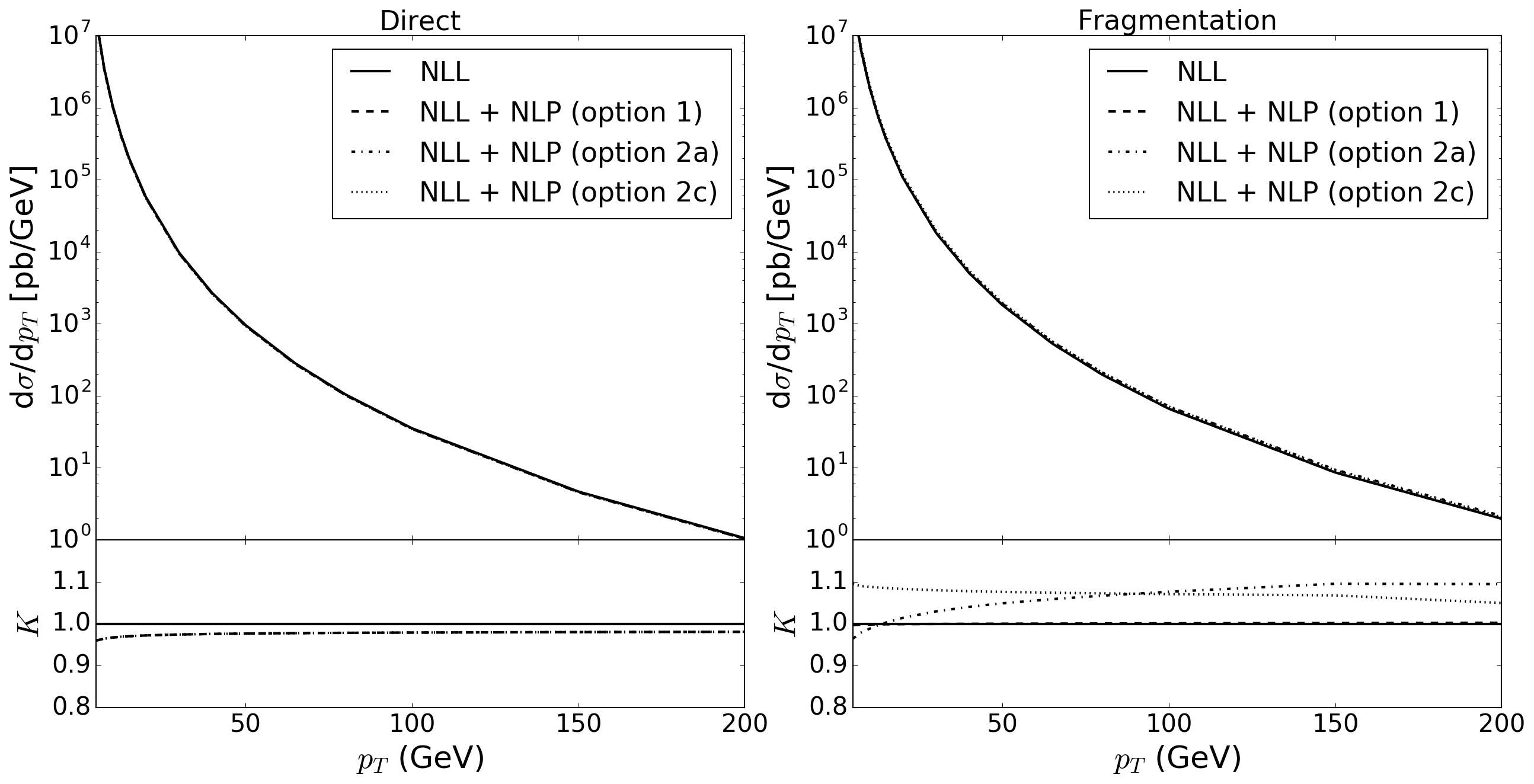}
  \caption{Joint-resummed $p_T$ distribution of the direct (left) and fragmentation (right) component for a central scale value of $\mu =  Q$ for the various options listed in section \ref{sec:fs}. 
  The bottom panels show the ratio $K$ with respect to the LP NLL resummed result.}
  \label{fig:FSNLPJR}
\end{figure}

The effect of the final state NLP terms on the joint-resummed $p_T$ distribution is shown in fig.~\ref{fig:FSNLPJR}. The initial state NLP terms are not included here. The left (right) figure shows the contribution to the direct (fragmentation) component. We observe a clear difference between the direct and fragmentation NLP results. The direct component is affected by final state NLP terms through the modified final state exponent for the unobserved parton. This yields a modest $ -5\%$ to $ -2\%$ difference with respect to the LP NLL result. Just as for LP terms, we observe that final state NLP corrections suppress the resummed distribution. Option 2a and 2c do not change the direct component as it contains no fragmentation function. \\
For the fragmentation component we observe that the ME approach hardly results in a correction to the NLL result. This is due to a cancellation between the exponents $E^{\rm PT}_c$ and $F_d$ in \eqref{eq:11}. The final state LL NLP diagonal evolution of the fragmentation function (option 2a) does not coincide with the ME approach, in contrast to the initial state case. This can be understood, as the fragmentation component has a different $N$ dependence than the resummed exponent (it depends on $2N+1$ rather than $N$, see   \eqref{eq:thresholdfrag}) and this creates spurious $1/N$ terms. Option 2a modifies the NLL differential cross-section by about $10\%$ for large $p_T$. If one allows a full evolution of the fragmentation function (option 2c), the differential cross-section is modified by $10\%$ for small $p_T$ values and $5\%$ at very large $p_T$. 

\subsection{Combined result}
\begin{figure}
  \centering
  \includegraphics[trim=0 0 0 0,clip,width=\textwidth]{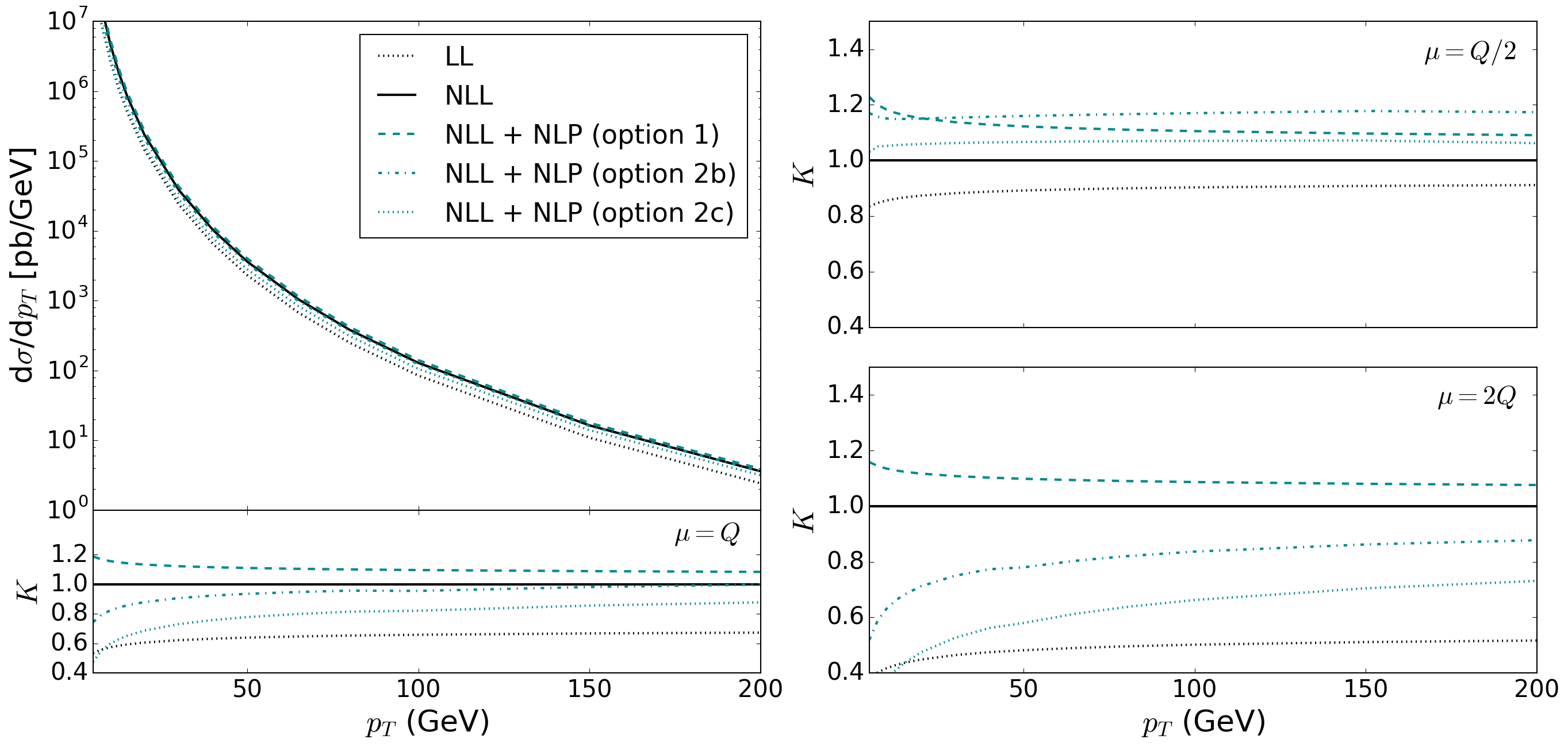}
  \caption{Combined joint-resummed $p_T$ distribution for LP resummation (black) and inclusion of NLP effects for both the initial and final state (blue). The bottom pannel and the plots on the right show the ratio $K$ with respect to the LP NLL resummed result for three choices of scale: $\mu =   Q$ (left), $\mu =  Q/2$ (top right) and $\mu =  2 Q$ (lower right).}
  \label{fig:ISandFSNLPJR}
\end{figure}

\begin{figure}
  \centering
  \includegraphics[trim=0 0 0 0,clip,width=\textwidth]{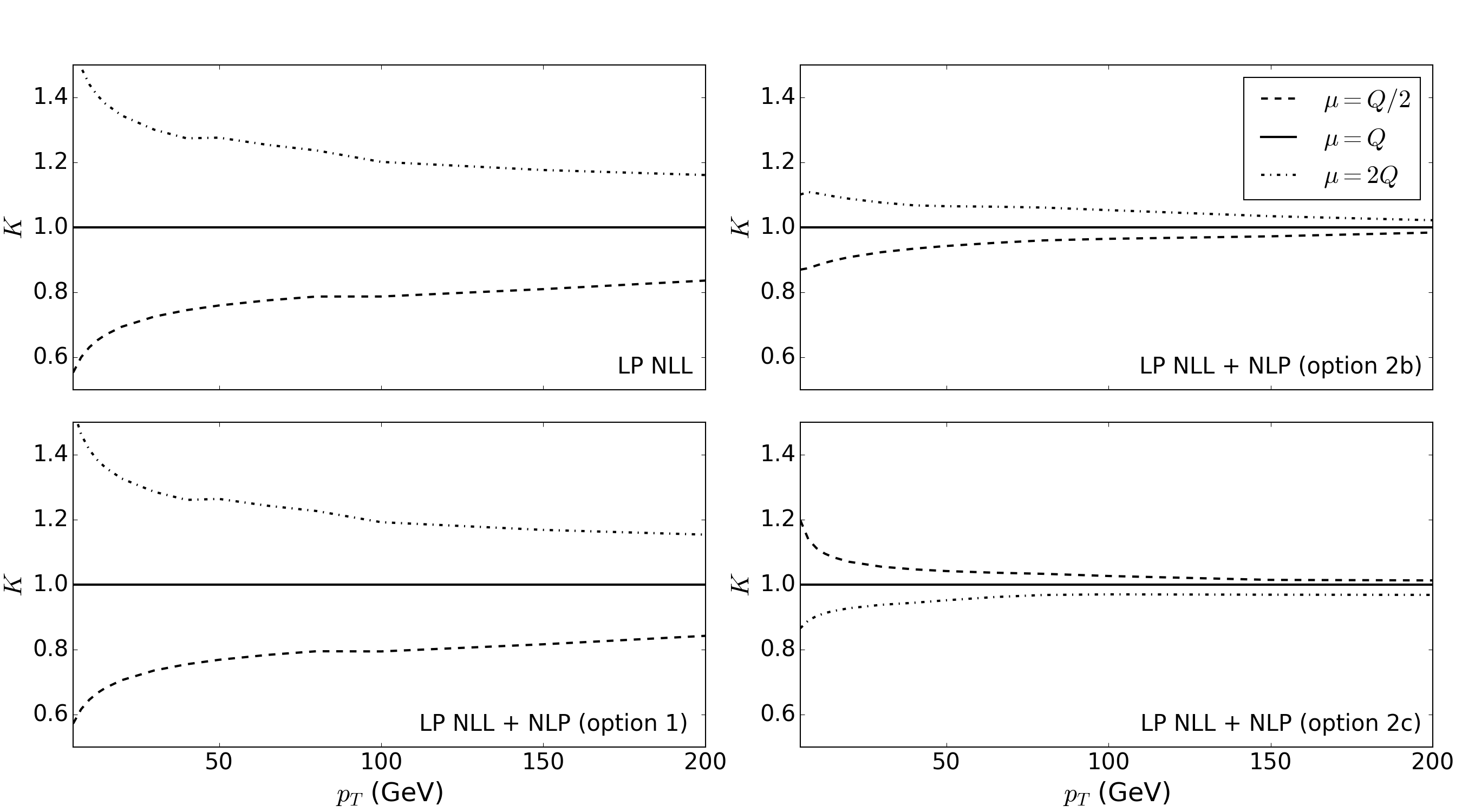}
  \caption{The ratio $K$ of the combined joint-resummed $p_T$ distribution for $\mu = Q/2$ (dashed) and $\mu =  2Q$ (dash-dotted) to $\mu=Q$ (solid). Four levels of accuracy are shown: LP NLL (top left), option 1 for both the initial and final state (lower left), option 2b for the initial state and option 2a for the final state (top right) and option 2c (lower left).}
  \label{fig:bothcomp}
\end{figure}

The result after the inclusion of both the initial and final state NLP terms can be seen in fig.~\ref{fig:ISandFSNLPJR}, where we show the combined joint-resummed $p_T$ distribution for three scales. We observe that the ratio of option 1 to the LP NLL resummed result amounts to a $10-20\%$ correction and is robust under scale variations. On the other hand, the ratios that are obtained with the evolution approaches (option 2b and 2c) are highly scale dependent. In fig.~\ref{fig:bothcomp} we can observe that the scale dependence of these ratios is in fact caused by the LP NLL joint-resummed $p_T$ distribution. The distributions obtained with option 2b (and 2c) are robust under variations of scale. We see that the NLP result obtained via option 1 also shows a large scale dependence. The scale dependence of the LP NLL result is therefore inherited by the result obtained using option 1, hence in their ratios shown in fig~\ref{fig:ISandFSNLPJR} we do not observe the large scale dependence. It is interesting to observe a significant decrease in scale dependence only after including the off-diagonal components of the splitting functions (option 2b for the initial state). The notable scale dependence at LP NLL for the joint-resummed $p_T$ distribution is connected to the indicated size contribution of the NLP corrections. More specifically, a diminishing of the LP scale dependence requires higher power terms, which is indeed what we find here. \\

\begin{figure}
  \centering
  \includegraphics[trim=0 0 0 0,clip,width=\textwidth]{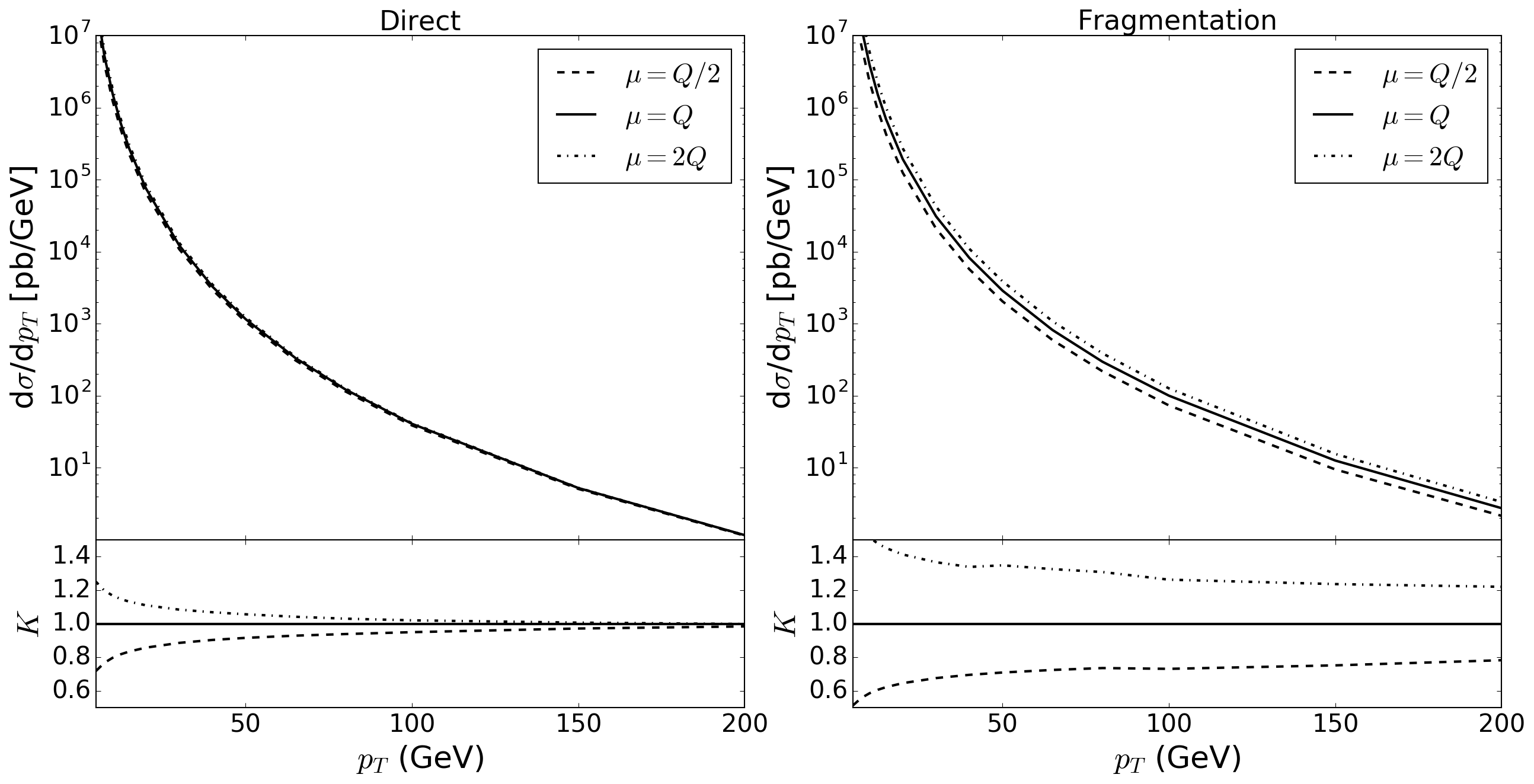}
  \caption{The scale variations of the  joint-resummed direct (left) and fragmentation (right) $p_T$ distribution, where option 1 of sections \ref{sec:is} and \ref{sec:fs} is used to include the NLP terms. The bottom panels show the ratio $K$ with respect to the scale choice of $\mu =  Q$. }
  \label{fig:scalevar}
\end{figure}

\begin{figure}
  \centering
  \includegraphics[trim=0 0 0 0,clip,width=\textwidth]{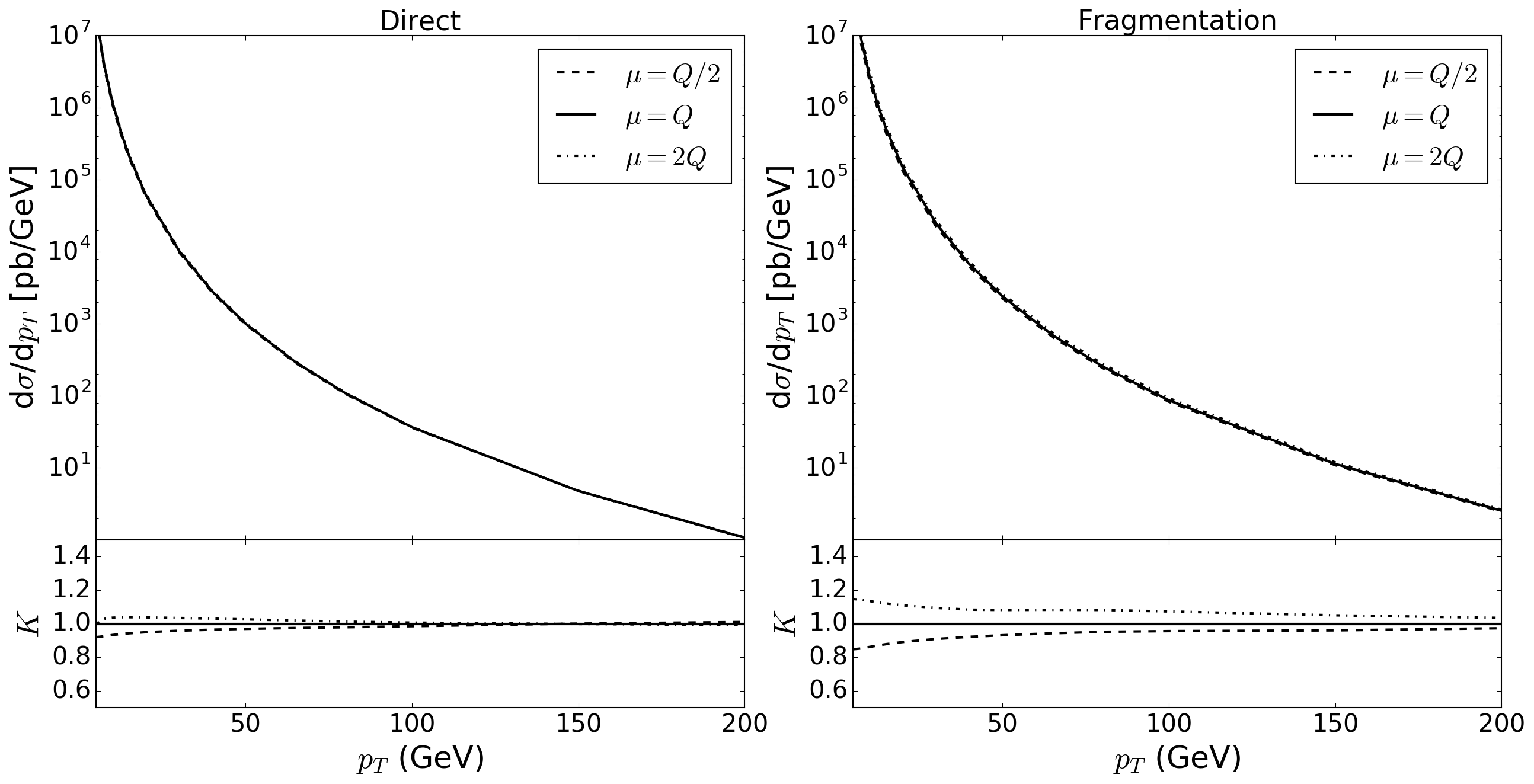}
  \caption{The scale variations of the  joint-resummed direct (left) and fragmentation (right) $p_T$ distribution, where option 2a of section \ref{sec:is} and 2b of section \ref{sec:fs} is used to include the NLP terms. The bottom panels show the ratio $K$ with respect to the scale choice of $\mu =  Q$.}
  \label{fig:scalevarevol}
\end{figure}

The origin of the large scale dependence observed for option 1 can be seen in fig.~\ref{fig:scalevar}. The direct component does not show a large dependence on the scale, and for large $p_T$ values the dependence even vanishes. The fragmentation component however depends considerably on the scale for option 1. At large $p_T$ it shows a $\pm 20\%$ scale variation, which at small $p_T$ becomes $\pm 60\%$. As noted before, the scale dependence was already present in the LP NLL result and the inclusion of only next-to-soft gluon emission does not decrease it. As can be seen in fig.~\ref{fig:scalevarevol}, using option 2b for the initial and option 2a for the final state NLP terms, the scale dependence of the fragmentation component is reduced tremendously. The scale dependence nearly vanishes for $p_T$ above $10$ GeV, and grows to about $15\%$ at very low $p_T$. This suggests that the emission of soft quarks plays an important role in stabilizing the dependence of differential distributions on scale variations. \\

\clearpage

\section{Conclusions}
\label{sec:conclusions}
In this paper we investigated for prompt photon production the difference between two previously proposed methods to include dominant $\ln{N}/N$ effects for joint resummation at NLL accuracy. These next-to-leading-power effects can be included either by modifying the resummation exponentials \cite{Basu:2007nu, Kramer:1996iq}, or by extending the evolution of the PDFs and fragmentation functions to $N$-dependent scales \cite{Kulesza:2002rh}. Both approaches only modify the initial or final state exponents, which capture all effects that are of collinear origin. We do not include non-collinear next-to-soft emission of partons, shifts of the hard scattering kinematics as a result of next-to-soft gluon emission, or the effects of the NLP phase space \cite{vanBeekveld:2019prq}.
By using the modified exponent approach for the initial state, one only accounts for the (next-to-)soft collinear gluon LL NLP contributions. This approach agrees exactly with the evolution approach if one only includes the diagonal splitting functions truncated to the right logarithmic order. There is an $\mathcal{O}(10\%)$ difference between the modified exponent approach and the evolution approach for the final state. This is due to the fragmentation function depending on $2N+1$ rather than $N$ (see \eqref{eq:thresholdfrag}). The correction obtained by including (next-to-)soft collinear gluon emission for both the initial and final state is about $10\%$ for high $p_T$ values and $20\%$ for low $p_T$.
\\
The off-diagonal components of the splitting functions represent the process where an initial state gluon splits into a quark-antiquark pair, or an initial state (anti-)quark splits into a gluon (anti-) quark pair. These contributions result in a NLP LL correction. The correction that is obtained by including both gluon and quark emission at NLP LL accuracy diminishes at central scale, which is caused by a sign difference between the diagonal and off-diagonal part of the splitting functions. However, at different scale choices, the correction can grow to more than $40\%$ for small $p_T$ values and $20\%$ for large $p_T$. This is due to the fact that the LP NLL resummed result is heavily scale dependent, and this dependence is greatly reduced by the inclusion of soft quark emission. This shows that soft quark emission is relevant at next-to-leading power \cite{vanBeekveld:2019prq, Moult:2019mog} and deserves further attention.

\subsection*{Acknowledgments}
This article is based upon work from COST Action CA16201 PARTICLEFACE
supported by COST (European Cooperation in Science and Technology).
WB, EL and MvB acknowledge support from the Dutch NWO-I program 156, "Higgs as Probe and Portal". AM would like to thank Nikhef, Amsterdam, ICTP, Trieste and Theory Division, CERN, Geneva  for their 
kind hospitality during the progress of this work. PM was supported by the Bundesministerium f\"{u}r Bildung und Forschung (BMBF). We wish to thank Benjamin Knorr and Werner Vogelsang for useful discussions. 

\appendix

\section{Exponents}
\label{app:exponents}

Here we list the exponents used in this paper.
The initial state exponents for the LL and NLL case without inclusion of the $\ln N/N$--terms are given by
\begin{eqnarray}
\label{eq:14a1}
h_a^{(0)} (\lambda) & =& \frac{A_a^{(1)}}{2\pi b_0^2}\Big[2\lambda + (1-2\lambda)\ln(1-2\lambda) \Big] \\
\label{eq:14a2}
h_a^{(1)} (\lambda,Q,\mu_F,\mu)& =& \frac{1}{2\pi
  b_0}\left(-\frac{A^{(2)}_a}{\pi b_0} + A^{(1)}_a \ln\Big(\frac{Q^2}{\mu^2} \Big) \right)\Big[2\lambda + \ln(1-2\lambda) \Big] \nonumber \\
            &  +& \frac{A^{(1)}_a b_1}{2\pi b_0^3}\Big[2\lambda + \ln(1-2\lambda)+\frac{1}{2}\ln^2(1-2\lambda) \Big] \nonumber \\
            &  -& \frac{A^{(1)}_a}{\pi b_0}\lambda \ln\Big(\frac{Q^2}{\mu_F^2} \Big) \; ,
\end{eqnarray}
where $\lambda = b_0 \alpha_s \ln \bar{N}$ and $\alpha_s\equiv\alpha_s(\mu^2)$. The function $h'$ that is added to account for the $\ln N/N$ terms is
\begin{eqnarray}
\label{eq:hprimeNLP}
h'_a(\lambda,\alpha_s) = -\frac{A^{(1)}_a}{2\pi b_0} \frac{\left[\ln(1-2\lambda)\right]}{N}.
\end{eqnarray}
For the case when evolving the parton distribution functions down to $Q/\bar{N}$, the initial state exponents are given by:
\begin{eqnarray}
\label{eq:14b}
\hat{h}_a^{(0)} (\lambda) &=& \frac{A_a^{(1)}}{2\pi b_0^2}
\left[ 2 \lambda + \ln(1-2 \lambda) \right]\, ,\\
\hat{h}_a^{(1)} (\lambda,Q,\mu) &=&
\frac{A_a^{(1)} b_1}{2\pi b_0^3} \left[ \frac{1}{2} \ln^2 (1-2 \lambda) +
\frac{2\lambda + \ln(1-2 \lambda)}{1-2\lambda} \right] + 
\frac{B_a^{(1)}}{\pi b_0}  \ln(1-2 \lambda) \nonumber \\
&+& \frac{1}{2\pi b_0} \left[ A_a^{(1)}\ln \left( \frac{Q^2}{\mu^2} \right)
-\frac{A_a^{(2)}}{\pi b_0}\right] \;
\left[ \frac{2 \lambda}{1-2\lambda}+ \ln(1-2 \lambda) \right] \; .
\end{eqnarray}
In the case of joint resummation, the recoil correction term that is added in both cases is given by
\begin{eqnarray}
h^{(1)}_{a, {\rm recoil}}(\lambda,\alpha_s) =A^{(1)}_a\frac{\alpha_s}{2\pi}\left(\frac{\zeta(2)}{1-2\lambda}+\frac{\ln\bar{N}}{N}\right).
\end{eqnarray}
The coefficients $A^{(n)}_a$ are given by: 
\begin{equation}
\label{eq:37}
    A_a^{(1)} = C_a, \qquad
    A_a^{(2)} = 
\frac{C_a}{2} \left[C_A\Bigg(\frac{67}{18}-\frac{\pi^2}{6}\Bigg)-\frac{10}{9}T_R n_f \right]
\end{equation}
with $C_q = C_F$ and $C_g = C_A$. Also we have
\begin{equation}
  \label{eq:20}
  B_q^{(1)} = -\frac{3}{4}C_F, \qquad
  B_g^{(1)} = -\pi b_0 \,.
\end{equation}
In these equations
\begin{eqnarray}
b_0 &=& \frac{11 C_A - 4 T_R n_f}{12 \pi}\;\;\;\; , \;\;\;\;\;
b_1 \;=\; \frac{17 C_A^2-10 C_A T_R n_f-6 C_F T_R n_f}{24 \pi^2}\; 
\end{eqnarray}
and $T_R = 1/2$, $C_A = 3$ and $C_F = \frac{4}{3}$. The number of active flavors is denoted by $n_f$.\\
The final state exponents involve the functions
\begin{equation}
\label{eq:39}
f^{(0)}_a(\lambda)= -\frac{A^{(1)}_a}{2\pi b_0^2}
          [(1-2 \lambda)\ln(1-2 \lambda) -2(1- \lambda)\ln(1- \lambda)]
\end{equation}   
\begin{eqnarray}
\label{eq:7}
f^{(1)}_a(\lambda,Q,\mu)=& -&\frac{A^{(1)}_a b_1}{2\pi b_0^3 }
            [\ln(1-2 \lambda) -2\ln(1- \lambda)
         +\frac{1}{2}\ln^2(1-2 \lambda)-\ln^2(1-\lambda)] \nonumber \\
        &+&\frac{B^{(1)}_a}{ \pi b_0 }\ln(1-\lambda)
      -\frac{A^{(2)}_a}{2\pi^2 b_0^2}[2 \ln(1-\lambda) - \ln(1-2\lambda)] \nonumber \\
 &+&\frac{A^{(1)}_a}{2\pi b_0}[2\ln(1- \lambda)- \ln(1-2\lambda)]\ln\frac{Q^2}{\mu^2}.
\end{eqnarray}
Our definition of $Q^2$ differs by a factor of $\frac{1}{2}$ from the definition of ref. \cite{Catani:1998tm}. To be able to compare our results to the first threshold resummation result that was computed in this reference, we add the term \cite{Laenen:2000ht}
\begin{eqnarray*}
-\frac{A^{(1)}_a}{\pi b_0}\ln(2)[\ln(1- \lambda)- \ln(1-2\lambda)]
\end{eqnarray*}
to \eqref{eq:7}. This term originates from setting $Q^2 \rightarrow 2 Q^2$ in the initial state and final state exponents. The terms that are added to account for the NLP effects of the final state parton read
\begin{equation}
\label{eq:25}
f^\prime_a(\lambda,\alpha_s) 
       =\frac{A^{(1)}_a}{ 2 \pi b_0}
 \frac{\left[\ln(1-2 \lambda)-\ln(1- \lambda)\right]}{N}\,,
\end{equation}
The wide-angle soft radiation exponents in   \eqref{eq:profile}  are given by
\begin{equation}
  \label{eq:8}
g^{(1)}_{q\bar{q}g}(\lambda,\mu) = -\frac{C_A}{\pi b_0}\ln (1-2\lambda) \ln 2,\qquad
g^{(1)}_{qgq}(\lambda,\mu) = -\frac{C_F}{\pi b_0}\ln (1-2\lambda) \ln 2.
\end{equation}
The explicit forms of $C_{\delta}^{(ab\to\gamma d)}(\alpha_s,\tilde{x}_T^2)$ of   \eqref{eq:direct} can be found in \cite{Catani:1998tm,Catani:1999hs}. The hard scattering matching coefficients $C_{\delta}^{(ab\to c d)}(\alpha_s,\tilde{x}_T^2)$ in   \eqref{eq:frag} are calculated by expanding the resummed cross-section to $\mathcal{O}(\alpha_s^3)$ and matching to the fixed order NLO result in ref. \cite{Aversa:1988vb}.

\section{Comparison with NLO}
\label{app:NLOcomparison}

In this appendix we compare the results of the modified exponent and the extended evolution approaches with the fixed order NLO calculation \cite{Aurenche:1983ws, Aurenche:1987fs,Gordon:1993qc}. The NLO cross section can be written in the following form
\begin{eqnarray}
v(1-v)s\frac{{\rm d}\sigma}{{\rm d}{v}{\rm d}w} &=&  s \left(\frac{4\pi\mu^2}{s}\right)^{2\varepsilon}\frac{v^2 (1-v)\left(v^2(1-v)w(1-w)\right)^{-\varepsilon}}{2(4\pi)^{4}\Gamma(1-2\varepsilon)} \nonumber \\
&& \hspace{1cm} \int_0^{\pi}{\rm d}\theta_1 \left(\sin\theta_1\right)^{1-2\varepsilon}\int_0^{\pi}{\rm d}\theta_2 \left(\sin\theta_2\right)^{-2\varepsilon}\langle|\mathcal{A}(v,w)|^2\rangle. 
\end{eqnarray}
with $ d=4-2\varepsilon$  the number of dimensions, and $\langle|\mathcal{A}|^2\rangle$ indicates the squared amplitude summed (averaged) over final (initial) state colours and spins. The azimuthal angles of the final state unobserved partons are given by $\theta_1$ and $\theta_2$ in the CM frame and the variables $v$ and $w$ are defined as 
\begin{eqnarray}
(p_a - p_{\gamma})^2 \equiv -svw \nonumber \,,\\
(p_b - p_{\gamma})^2  \equiv s(v-1) \nonumber\,, \\ 
(p_a+p_b-p_{\gamma})^2 \equiv sv(1-w).
\end{eqnarray}
The phase space measure for the collinear counter term is that of a two-body final state, and is given by 
\begin{eqnarray}
{\rm d}\Phi_2 = \left(\frac{4\pi\mu^2}{s}\right)^{\varepsilon}\frac{\left(v(1-v)\right)^{-\varepsilon}}{8\pi \Gamma(1-\varepsilon)}\int {\rm d} v\int {\rm d}w \,\delta(1-w).
\end{eqnarray}
The full process at NLO consists of a three-body body final state, and the difference between its phase space measure and with that of the counterterm leads to a class of LP and NLP logarithms. After performing mass factorization and setting $\mu = Q$, the cross section can be written as
\begin{eqnarray}
\left(\frac{\alpha \alpha_s^2 Q_q^2}{v (1-v) s}\right)^{-1} \frac{{\rm d} \sigma_{q\bar{q}}}{{\rm d} v {\rm d} w} &=& c_1'(v)  \frac{1}{\varepsilon}\frac{1}{(1-w)}_+ \left[-\varepsilon\ln(v)-\varepsilon\ln(1-w)\right] \nonumber \\
&& + c_2'(v) \frac{1}{\varepsilon}\left[-\varepsilon\ln(1-w)\right]  + c_3'(v) \frac{1}{(1-w)}_++ c_4'(v) \ln(1-w)  +\dots \nonumber \\
&\equiv& c_1(v) + c_2(v)+ c_3(v)+ c_4(v) +\dots\,,
\label{eq:struc}
\end{eqnarray}
where we have brought a common factor to the left and the ellipsis indicates contributions that do not correspond to LP or NLP logarithms. To compare with the resummation exponents, we transform these coefficients to $N$ space. 
The Mellin moment is taken with respect to $x_T^2 = 4vw(1-v)$,
\begin{eqnarray}
\sigma(N) \equiv \int^1_0{\rm d} x_T^2 \, (x_T^2)^{N-1} \, \left[\frac{p_T^3{\rm d}\sigma(p_T)}{{\rm d}p_T}\right] = \frac{p_T^2}{8x_a x_b S} \int^1_0{\rm d} v \int^1_0 {\rm d} w \,\,\,(4v(1-v)w)^{N}\, \frac{s{\rm d}\sigma(v,w)}{{\rm d}v {\rm d}w},
\end{eqnarray}
where the integral over $x_T^2$ is for convenience expressed  \cite{Catani:1998tm} as two integrals over $v$ and $w$. We extracted an overall factor of $\frac{p_T^2}{8x_a x_b S}$ to match the equations we use for the resummed differential distributions (e.g. \eqref{eq:thresholddirect}).  Note that with this choice our $N$-space expressions differ from those presented in \cite{Catani:1998tm}. We now proceed to determine the various origins of the coefficients in \eqref{eq:struc}. We 
only consider the two channels in which the photon can be produced directly, see \eqref{eq:parton-proc}. The fragmentation channels can be analyzed analogously, hence we do not present them here.  

In \cite{vanBeekveld:2019prq} we have performed an NLP analysis of the prompt photon production process at NLO. We express these results here in the form of \eqref{eq:struc} and find for the $q\bar{q}\rightarrow g g \gamma$ channel
\begin{eqnarray}
c^{q\bar{q}}_1(v) &=&\left[\left(\frac{\ln(1-w)}{(1-w)}\right)_++\frac{\ln(v)}{(1-w)}_+\right]\times\left\{4\frac{C_F^2}{C_A} T_{q\bar{q}}\right\}\,, \\
c^{q\bar{q}}_2(v) &=& \ln(1-w)\times \left\{\frac{2C_F^2}{C_A}\frac{1-2v}{1-v}\right\}   \,,\\
c_3^{q\bar{q}}(v)&=&  \frac{1}{(1-w)}_+\times\Bigg\{4\frac{C_F^2}{C_A}T_{q\bar{q}}\left(\ln(1-w)-\ln(1-v)\right) \\
&& \hspace{2.5 cm}-2C_FT_{q\bar{q}}\left(\ln(1-w)-\ln(1-v)\right)-\frac{11 C_F}{6}T_{q\bar{q}} +\frac{C_Fn_f}{3C_A}T_{q\bar{q}}\Bigg\} \,, \nonumber\\
c_4^{q\bar{q}}(v) &=&  \ln(1-w)\times\left\{\left(\frac{2C_F^2}{C_A}-C_F\right)\frac{1-2v}{1-v}\right\}\,,
\end{eqnarray}
where 
$T_{q\bar{q}} = 2(v-1)v+1$. The relevant $v$ and $w$ integrals in the Mellin transformation are as follows
\begin{eqnarray}
\int_0^1{\rm d}w \,w^{N}\left(\frac{\ln(1-w)}{1-w}\right)_+ &\simeq& \frac{1}{2}\left(\ln^2\bar{N}+\frac{\ln\bar{N}}{N}\right)+\mathcal{O}\left(1\right) , \label{eq:ln1mw}\\
\int_0^1{\rm d}w \,w^{N}\frac{1}{(1-w)}_+ &\simeq& -\ln\bar{N}+\mathcal{O}\left(\frac{1}{N}\right) ,\\
\int_0^1{\rm d}w \,w^{N}\ln(1-w) &\simeq& -\frac{\ln\bar{N}}{N}+\mathcal{O}\left(\frac{\ln\bar{N}}{N^2}\right)   ,\\
4\int_0^1{\rm d}v \,(4v(1-v))^{N-1}T_{q\bar{q}}  &=& \frac{\Gamma(1/2)\Gamma(N)(N+1)}{\Gamma \left(N+3/2\right)}\equiv T_{q\bar{q}}(N), \\
4\int_0^1{\rm d}v \,(4v(1-v))^{N-1} \, T_{q\bar{q}}\, \ln(v)  &=&
4\int_0^1{\rm d}v \,(4v(1-v))^{N-1}\,T_{q\bar{q}}\,\ln(1-v)  \nonumber\\
&\simeq& T_{q\bar{q}}(N)\left(-\ln(2)-\frac{1}{4N}+\mathcal{O}\left(\frac{1}{N^2}\right)\right).
\label{eq:ln1mv}
\end{eqnarray}
The $N$-space results then read
\begin{eqnarray}
c_1^{q\bar{q}}(N)&=& T_{q\bar{q}}(N)\frac{C_F^2}{C_A}\Bigg[2\bigg(\ln^2\bar{N}+\underbrace{\frac{\ln\bar{N}}{N}}_{\text{1 \& 2}}\bigg)+\bigg(4\ln(2)\ln\bar{N}+\underbrace{\frac{\ln\bar{N}}{N}}_{\text{PS}}\bigg)+\mathcal{O}\left(1\right)\Bigg] \,,\\
c_2^{q\bar{q}}(N) &=&
\mathcal{O}\left(\frac{\ln\bar{N}}{N^2}\right) ,\\
c_3^{q\bar{q}}(N) &=& T_{q\bar{q}}(N) \frac{C_F}{C_A}\Bigg[(2C_F-C_A)\Bigg(\Big(\ln^2\bar{N}+\underbrace{\frac{\ln\bar{N}}{N}}_{\text{1 \& 2}}\Big)-\bigg(4\ln(2)\ln\bar{N}+\underbrace{\frac{\ln\bar{N}}{N}}_{\text{PS}}\bigg)\Bigg) \nonumber\\
&&\hspace{6cm}+\frac{11C_A-2n_f}{6}\ln\bar{N}+\mathcal{O}\left(1\right)\Bigg]  \,, \\
c_4^{q\bar{q}}(N) &=& \mathcal{O}\left(\frac{\ln\bar{N}}{N^2}\right).
\end{eqnarray}
Underneath the LL NLP terms we have indicated their origin. To compare these results with the threshold resummation formulae we expand \eqref{eq:hprimeNLP} to NLO. The NLP logarithms stemming from \eqref{eq:ln1mw} are captured by every option (indicated with 1 \& 2). Those stemming from \eqref{eq:ln1mv}, indicated with PS, stem from the difference between the two-body and three-body phase space measure in the mass factorization procedure. 
They are omitted in the numerical analysis of this paper,
as they do not have the factorized form needed for the resummation formulae used. Note that $c_2$ and $c_4$ only contribute beyond NLP. 

\noindent The other direct production channel, $q g \rightarrow q g \gamma$, has two sources of NLP logarithms: soft gluon and soft quark emission. The soft gluon contribution after performing mass factorization reads
\begin{eqnarray}
c^{qg}_{1,g}(v) &=&\left[\left(\frac{\ln(1-w)}{(1-w)}\right)_++\frac{\ln(v)}{(1-w)}_+\right]\times \left\{\left(1+\frac{C_F}{C_A}\right) T_{qg}\right\} \,, \\
c^{qg}_{2,g}(v) &=&\ln(1-w)\times \left\{\frac{(v-2)v^3}{1-v}\right\} \,, \\
c^{qg}_{3,g}(v) &=& \frac{1}{(1-w)_+}\times\Big\{T_{qg}\left(\ln(1-w)-\ln(1-v)-\frac{C_F}{C_A}\right)\Big\}\,, \\
c^{qg}_{4,g}(v) &=& \ln(1-w)\times\left\{\frac{1}{2}\frac{(v-2)v^3}{1-v}\right\},
\end{eqnarray}
where $T_{qg} = v((v-2)v+2)$. The Mellin moments are given by
\begin{eqnarray} 
c^{qg}_{1,g}(N) &=& T_{qg}(N)\left(1+\frac{C_F}{C_A}\right)\Bigg[\frac{1}{2}\bigg(\ln^2\bar{N}+\underbrace{\frac{\ln\bar{N}}{N}}_{\text{1 \& 2}}\bigg) +\bigg(\ln(2)\ln\bar{N}-\underbrace{\frac{1}{20}\frac{\ln\bar{N}}{N}}_{\text{PS}}\bigg)+\mathcal{O}(1)\Bigg], \\
c^{qg}_{2,g}(N) &=& T_{qg}(N)\underbrace{\frac{3}{5}\frac{\ln\bar{N}}{ N}}_{\text{Low}}\,, \\
c^{qg}_{3,g}(N) &=&   T_{qg}(N)\Bigg[\frac{1}{2}\bigg(\ln^2\bar{N}+\underbrace{\frac{\ln\bar{N}}{N}}_{\text{1 \& 2}}\bigg) -\bigg(\ln(2)\ln\bar{N}+\underbrace{\frac{11}{20}\frac{\ln\bar{N}}{N}}_{\text{PS}}\bigg)\nonumber \\
&& \hspace{8cm} +\frac{C_F}{C_A}\ln\bar{N}+\mathcal{O}(1)\Bigg]\,,  \\
c^{qg}_{4,g}(v) &=& T_{qg}(N)\underbrace{\frac{3}{10}\frac{\ln\bar{N}}{ N}}_{\text{Low}}\,,
\end{eqnarray}
where we have used the following expressions for the $v$ integrals
\begin{eqnarray}
4\int_0^1{\rm d}v \,(4v(1-v))^{N-1}\,T_{qg}  &=& \frac{\Gamma(1/2)\Gamma (N)(5N+2)}{4\,\Gamma\left(N+3/2\right)}\equiv T_{qg}(N) , \\
4\int_0^1{\rm d}v \,(4v(1-v))^{N-1} \, T_{qg}\, \ln(v)  &\simeq& T_{qg}(N)\left(-\ln(2)+\frac{1}{20}\frac{1}{N}+\mathcal{O}\left(\frac{1}{N^2}\right)\right) \label{eq:lnv}, \\
4\int_0^1{\rm d}v \,(4v(1-v))^{N-1}\,T_{qg}\,\ln(1-v) 
&\simeq& T_{qg}(N)\left(-\ln(2)-\frac{11}{20}\frac{1}{N}+\mathcal{O}\left(\frac{1}{N^2}\right)\right).\label{eq:ln1mv2}
\end{eqnarray}
As in the $q\bar{q}\rightarrow g g\gamma $ channel, the NLP logarithms that originate from \eqref{eq:lnv} and \eqref{eq:ln1mv2} are not captured. The origin of the missing NLP logarithms in $c_2$ and $c_4$ is the expansion of the hard scattering matrix element (Low), associated with Low's theorem \cite{Low:1958sn, Burnett:1967km, DelDuca:1990gz, DelDuca:2017twk, vanBeekveld:2019prq}. To capture these logarithms, one would need to modify the hard scattering functions and allow their momentum dependence to deviate from exact threshold. \\
Next we show the mass factorized soft quark contributions
\begin{eqnarray}
c^{qg}_{1,q}(v) &=& 0 \,, \\
c^{qg}_{2,q}(v) &=& \ln(1-w)\times\left\{\frac{1}{2}\frac{v^2(v^2+1)}{1-v} + \frac{1}{2}\frac{C_F}{C_A}\frac{v(v(v((v-2)v+4)-4)+2)}{1-v}\right\}\,, \\
c^{qg}_{3,q}(v) &=& \frac{1}{(1-w)}_+\times\left\{\frac{1}{4}\frac{C_F}{C_A} T_{qg}\right\} \,, \\
c^{qg}_{4,q}(v) &=& \ln(1-w)\times\left\{\frac{1}{2}\frac{v^4}{1-v}+\frac{C_F}{C_A}v^3 \right\}. \label{eq:softquark}
\end{eqnarray}
Two separate origins contribute to the coefficient $c^{qg}_{2,q}(v)$: an initial state splitting of a gluon into a quark-anti-quark pair, and the fragmentation of a quark into a quark-photon pair. The contributions read
\begin{eqnarray}
{\text{Initial state contribution: }}&& c^{qg}_{2,q,{\rm IS}}(v) = \ln(1-w)\times\left\{\frac{1}{2}\frac{C_F}{C_A}\frac{v T_{q\bar{q}}}{1-v}\right\}\,, \\
{\text{Fragmentation contribution: }}&& c^{qg}_{2,q,{\rm FS}}(v) = \ln(1-w) \\
&& \hspace{3cm} \times\left\{\frac{1}{2}\frac{C_F}{C_A}v(1-v)(v^2+1) + \frac{1}{2}\frac{v^2(v^2+1)}{1-v}\right\}. \nonumber
\end{eqnarray}
For the initial state contribution, a gluon splits into a quark and an anti-quark, where the latter collides with the other initial state quark in the hard scattering process $q\bar{q}\rightarrow g \gamma$. The $N$-space initial state contribution reads
\begin{eqnarray}
c^{qg}_{2,q,{\rm IS}}(N) = -\frac{1}{2}\frac{C_F}{C_A}T_{q\bar{q}}(N)\underbrace{\frac{\ln\bar{N}}{N}}_{\text{2b \& 2c}} +\mathcal{O}\left(\frac{\ln\bar{N}}{N^2}\right).
\end{eqnarray}
which as indicated is part of option 2b and 2c, as can be seen if one expands the off-diagonal components of the splitting matrix. For the fragmentation contribution, the hard scattering process is $qg\rightarrow qg$. The $v$ integral in the Mellin moment is
\begin{eqnarray}
 4\int_0^1{\rm d}v \,(4v(1-v))^{N-1}\, \frac{c^{qg}_{2,q,{\rm FS}}(v)}{\ln(1-w)} &=& \frac{1}{16C_A} \frac{\Gamma(1/2)\Gamma(N-1)}{\Gamma(N+5/2)}((2N+3)(N(5N+11)+8)C_A \nonumber \\
&& \hspace{5cm}+(N-1)N(5N+8)C_F) \nonumber \\
& \equiv& \frac{1}{2}T_{qg\rightarrow qg}(N) \,. 
\end{eqnarray}
The $N$-space contribution of the fragmentation component after performing the $w$ integral reads
\begin{eqnarray}
c^{qg}_{2,q,{\rm FS}}(N) = -\frac{1}{2}T_{qg\rightarrow qg }(N) \underbrace{\frac{\ln\bar{N}}{N}}_{\text{2a \& 2c}} +\mathcal{O}\left(\frac{\ln\bar{N}}{N^2}\right) \,,
\end{eqnarray}
which is captured by option 2a and 2c as given in section~\ref{sec:fs},see \eqref{eq:plevol}. The other two terms of \eqref{eq:softquark} result in
\begin{eqnarray}
c^{qg}_{3,q}(N) &=&   T_{qg}(N)\frac{1}{4}\frac{C_F}{C_A}\left[-\ln\bar{N}+\mathcal{O}\left(\frac{1}{N}\right)\right]\,, \\
c^{qg}_{4,q}(N) &=& T_{qg}(N)\left(1+\frac{2 C_F}{C_A}\right)\Bigg[-\underbrace{\frac{1}{10}\frac{\ln\bar{N}}{N}}_{\text{Int}}+\mathcal{O}\left(\frac{1}{N^2}\right)\Bigg]\,,
\end{eqnarray}
The NLP logarithm (Int), which is not included in our numerical results, is a consequence of the interference between an initial and final state soft (non-collinear) quark.

\bibliographystyle{JHEP}
\bibliography{spires}

\providecommand{\href}[2]{#2}\begingroup\raggedright\begin{thebibliography}{10}

\bibitem{Sterman:1987aj}
G.~Sterman, {\it Summation of large corrections to short distance hadronic
  cross-sections},  {\em Nucl. Phys.} {\bf B281} (1987) 310.

\bibitem{Catani:1989ne}
S.~Catani and L.~Trentadue, {\it {Resummation of the QCD Perturbative Series
  for Hard Processes}},  {\em Nucl. Phys.} {\bf B327} (1989) 323.

\bibitem{Laenen:1998qw}
E.~Laenen, G.~Oderda, and G.~Sterman, {\it Resummation of threshold corrections
  for single particle inclusive cross-sections},  {\em Phys. Lett.} {\bf B438}
  (1998) 173--183, [\href{http://arxiv.org/abs/hep-ph/9806467}{{\tt
  hep-ph/9806467}}].

\bibitem{Catani:1998tm}
S.~Catani, M.~L. Mangano, and P.~Nason, {\it Sudakov resummation for prompt
  photon production in hadron collisions},  {\em JHEP} {\bf 9807} (1998) 024,
  [\href{http://arxiv.org/abs/hep-ph/9806484}{{\tt hep-ph/9806484}}].

\bibitem{Catani:1999hs}
S.~Catani, M.~L. Mangano, P.~Nason, C.~Oleari, and W.~Vogelsang, {\it Sudakov
  resummation effects in prompt photon hadroproduction},  {\em JHEP} {\bf 9903}
  (1999) 025, [\href{http://arxiv.org/abs/hep-ph/9903436}{{\tt
  hep-ph/9903436}}].

\bibitem{Kidonakis:1999hq}
N.~Kidonakis and J.~F. Owens, {\it Soft-gluon resummation and nnlo corrections
  for direct photon production},  {\em Phys. Rev.} {\bf D61} (2000) 094004,
  [\href{http://arxiv.org/abs/hep-ph/9912388}{{\tt hep-ph/9912388}}].

\bibitem{Sterman:2000pt}
G.~Sterman and W.~Vogelsang, {\it Threshold resummation and rapidity
  dependence},  {\em JHEP} {\bf 0102} (2001) 016,
  [\href{http://arxiv.org/abs/http://arXiv.org/abs/hep-ph/0011289}{{\tt
  http://arXiv.org/abs/hep-ph/0011289}}].

\bibitem{Sterman:2004yk}
G.~Sterman and W.~Vogelsang, {\it Recoil and power corrections in high-x(t)
  direct-photon production},  {\em Phys. Rev.} {\bf D71} (2005) 014013,
  [\href{http://arxiv.org/abs/hep-ph/0409234}{{\tt hep-ph/0409234}}].

\bibitem{deFlorian:2005wf}
D.~de~Florian and W.~Vogelsang, {\it Threshold resummation for the
  prompt-photon cross section revisited},  {\em Phys. Rev.} {\bf D72} (2005)
  014014, [\href{http://arxiv.org/abs/hep-ph/0506150}{{\tt hep-ph/0506150}}].

\bibitem{deFlorian:2005yj}
D.~de~Florian and W.~Vogelsang, {\it Threshold resummation for the inclusive
  hadron cross- section in p p collisions},  {\em Phys. Rev.} {\bf D71} (2005)
  114004, [\href{http://arxiv.org/abs/hep-ph/0501258}{{\tt hep-ph/0501258}}].

\bibitem{deFlorian:2013taa}
D.~de~Florian, M.~Pfeuffer, A.~Sch{\"a}fer, and W.~Vogelsang, {\it {Soft-gluon
  Resummation for High-pT Inclusive-Hadron Production at COMPASS}},  {\em Phys.
  Rev.} {\bf D88} (2013), no.~1 014024,
  [\href{http://arxiv.org/abs/1305.6468}{{\tt arXiv:1305.6468}}].

\bibitem{Bolzoni:2005xn}
P.~Bolzoni, S.~Forte, and G.~Ridolfi, {\it Renormalization group approach to
  sudakov resummation in prompt photon production},  {\em Nucl. Phys.} {\bf
  B731} (2005) 85--108, [\href{http://arxiv.org/abs/hep-ph/0504115}{{\tt
  hep-ph/0504115}}].

\bibitem{Becher:2009th}
T.~Becher and M.~D. Schwartz, {\it {Direct photon production with effective
  field theory}},  {\em JHEP} {\bf 1002} (2010) 040,
  [\href{http://arxiv.org/abs/0911.0681}{{\tt arXiv:0911.0681}}].

\bibitem{Becher:2012xr}
T.~Becher, C.~Lorentzen, and M.~D. Schwartz, {\it {Precision direct photon and
  W-boson spectra at high pT and comparison to LHC data}},  {\em Phys.Rev.}
  {\bf D86} (2012) 054026, [\href{http://arxiv.org/abs/1206.6115}{{\tt
  arXiv:1206.6115}}].

\bibitem{Hinderer:2018nkb}
P.~Hinderer, F.~Ringer, G.~Sterman, and W.~Vogelsang, {\it {Threshold
  Resummation at NNLL for single-particle Production in Hadronic Collisions}},
  \href{http://arxiv.org/abs/1812.00915}{{\tt arXiv:1812.00915}}.

\bibitem{Schwartz:2016olw}
M.~D. Schwartz, {\it {Precision direct photon spectra at high energy and
  comparison to the 8 TeV ATLAS data}},  {\em JHEP} {\bf 1609} (2016) 005,
  [\href{http://arxiv.org/abs/1606.02313}{{\tt arXiv:1606.02313}}].

\bibitem{Laenen:2000ij}
E.~Laenen, G.~Sterman, and W.~Vogelsang, {\it Recoil and threshold corrections
  in short-distance cross sections},  {\em Phys. Rev.} {\bf D63} (2001) 114018,
  [\href{http://arxiv.org/abs/http://arXiv.org/abs/hep-ph/0010080}{{\tt
  http://arXiv.org/abs/hep-ph/0010080}}].

\bibitem{Li:1998is}
H.~nan Li, {\it Unification of the k(t) and threshold resummations},  {\em
  Phys. Lett.} {\bf B454} (1999) 328,
  [\href{http://arxiv.org/abs/hep-ph/9812363}{{\tt hep-ph/9812363}}].

\bibitem{Laenen:2000de}
E.~Laenen, G.~Sterman, and W.~Vogelsang, {\it Higher-order qcd corrections in
  prompt photon production},  {\em Phys. Rev. Lett.} {\bf 84} (2000) 4296,
  [\href{http://arxiv.org/abs/hep-ph/0002078}{{\tt hep-ph/0002078}}].

\bibitem{Banfi:2004xa}
A.~Banfi and E.~Laenen, {\it {Joint resummation for heavy quark production}},
  {\em Phys.Rev.} {\bf D71} (2005) 034003,
  [\href{http://arxiv.org/abs/hep-ph/0411241}{{\tt hep-ph/0411241}}].

\bibitem{Bozzi:2007tea}
G.~Bozzi, B.~Fuks, and M.~Klasen, {\it {Joint resummation for slepton pair
  production at hadron colliders}},  {\em Nucl. Phys.} {\bf B794} (2008)
  46--60, [\href{http://arxiv.org/abs/0709.3057}{{\tt arXiv:0709.3057}}].

\bibitem{Fuks:2007gk}
B.~Fuks, M.~Klasen, F.~Ledroit, Q.~Li, and J.~Morel, {\it {Precision
  predictions for $Z^\prime$ - production at the CERN LHC: QCD matrix elements,
  parton showers, and joint resummation}},  {\em Nucl. Phys.} {\bf B797} (2008)
  322--339, [\href{http://arxiv.org/abs/0711.0749}{{\tt arXiv:0711.0749}}].

\bibitem{Debove:2011xj}
J.~Debove, B.~Fuks, and M.~Klasen, {\it {Joint Resummation for Gaugino Pair
  Production at Hadron Colliders}},  {\em Nucl. Phys.} {\bf B849} (2011)
  64--79, [\href{http://arxiv.org/abs/1102.4422}{{\tt arXiv:1102.4422}}].

\bibitem{Kulesza:2002rh}
A.~Kulesza, G.~Sterman, and W.~Vogelsang, {\it Joint resummation in electroweak
  boson production},  {\em Phys. Rev.} {\bf D66} (2002) 014011,
  [\href{http://arxiv.org/abs/hep-ph/0202251}{{\tt hep-ph/0202251}}].

\bibitem{Kulesza:2003wn}
A.~Kulesza, G.~Sterman, and W.~Vogelsang, {\it Joint resummation for higgs
  production},  {\em Phys. Rev.} {\bf D69} (2004) 014012,
  [\href{http://arxiv.org/abs/hep-ph/0309264}{{\tt hep-ph/0309264}}].

\bibitem{Marzani:2016smx}
S.~Marzani and V.~Theeuwes, {\it {Vector boson production in joint
  resummation}},  {\em JHEP} {\bf 02} (2017) 127,
  [\href{http://arxiv.org/abs/1612.01432}{{\tt arXiv:1612.01432}}].

\bibitem{Muselli:2017bad}
C.~Muselli, S.~Forte, and G.~Ridolfi, {\it {Combined threshold and transverse
  momentum resummation for inclusive observables}},  {\em JHEP} {\bf 03} (2017)
  106, [\href{http://arxiv.org/abs/1701.01464}{{\tt arXiv:1701.01464}}].

\bibitem{Lustermans:2016nvk}
G.~Lustermans, W.~J. Waalewijn, and L.~Zeune, {\it {Joint transverse momentum
  and threshold resummation beyond NLL}},  {\em Phys. Lett.} {\bf B762} (2016)
  447--454, [\href{http://arxiv.org/abs/1605.02740}{{\tt arXiv:1605.02740}}].

\bibitem{Campbell:2016lzl}
J.~M. Campbell, R.~K. Ellis, and C.~Williams, {\it {Direct photon production at
  NNLO}},  {\em Phys. Rev. Lett.} {\bf 118} (2017), no.~22 222001,
  [\href{http://arxiv.org/abs/1612.04333}{{\tt arXiv:1612.04333}}].

\bibitem{Kramer:1996iq}
M.~Kramer, E.~Laenen, and M.~Spira, {\it Soft gluon radiation in {H}iggs boson
  production at the {LHC}},  {\em Nucl. Phys.} {\bf B511} (1998) 523--549,
  [\href{http://arxiv.org/abs/hep-ph/9611272}{{\tt hep-ph/9611272}}].

\bibitem{Herzog:2014wja}
F.~Herzog and B.~Mistlberger, {\it {The Soft-Virtual Higgs Cross-section at
  N$^3$LO and the Convergence of the Threshold Expansion}},
  \href{http://arxiv.org/abs/1405.5685}{{\tt arXiv:1405.5685}}.

\bibitem{Low:1958sn}
F.~E. Low, {\it {Bremsstrahlung of very low-energy quanta in elementary
  particle collisions}},  {\em Phys. Rev.} {\bf 110} (1958) 974--977.

\bibitem{Burnett:1967km}
T.~H. Burnett and N.~M. Kroll, {\it {Extension of the low soft photon
  theorem}},  {\em Phys. Rev. Lett.} {\bf 20} (1968) 86.

\bibitem{DelDuca:1990gz}
V.~Del~Duca, {\it High-energy bremsstrahlung theorems for soft photons},  {\em
  Nucl. Phys.} {\bf B345} (1990) 369--388.

\bibitem{Laenen:2008gt}
E.~Laenen, G.~Stavenga, and C.~D. White, {\it {Path integral approach to
  eikonal and next-to-eikonal exponentiation}},  {\em JHEP} {\bf 0903} (2009)
  054, [\href{http://arxiv.org/abs/0811.2067}{{\tt arXiv:0811.2067}}].

\bibitem{Laenen:2010uz}
E.~Laenen, L.~Magnea, G.~Stavenga, and C.~D. White, {\it {Next-to-eikonal
  corrections to soft gluon radiation: a diagrammatic approach}},  {\em JHEP}
  {\bf 1101} (2011) 141, [\href{http://arxiv.org/abs/1010.1860}{{\tt
  arXiv:1010.1860}}].

\bibitem{Soar:2009yh}
G.~Soar, S.~Moch, J.~Vermaseren, and A.~Vogt, {\it {On Higgs-exchange DIS,
  physical evolution kernels and fourth-order splitting functions at large x}},
   {\em Nucl. Phys.} {\bf B832} (2010) 152--227,
  [\href{http://arxiv.org/abs/0912.0369}{{\tt arXiv:0912.0369}}].

\bibitem{Moch:2009hr}
S.~Moch and A.~Vogt, {\it {On non-singlet physical evolution kernels and
  large-x coefficient functions in perturbative QCD}},  {\em JHEP} {\bf 0911}
  (2009) 099, [\href{http://arxiv.org/abs/0909.2124}{{\tt arXiv:0909.2124}}].

\bibitem{Moch:2009mu}
S.~Moch and A.~Vogt, {\it {Threshold Resummation of the Structure Function
  F(L)}},  {\em JHEP} {\bf 0904} (2009) 081,
  [\href{http://arxiv.org/abs/0902.2342}{{\tt arXiv:0902.2342}}].

\bibitem{deFlorian:2014vta}
D.~de~Florian, J.~Mazzitelli, S.~Moch, and A.~Vogt, {\it {Approximate N$^{3}$LO
  Higgs-boson production cross section using physical-kernel constraints}},
  {\em JHEP} {\bf 1410} (2014) 176, [\href{http://arxiv.org/abs/1408.6277}{{\tt
  arXiv:1408.6277}}].

\bibitem{Presti:2014lqa}
N.~Lo~Presti, A.~Almasy, and A.~Vogt, {\it {Leading large-x logarithms of the
  quark \& gluon contributions to inclusive Higgs-boson and lepton-pair
  production}},  {\em Phys. Lett.} {\bf B737} (2014) 120--123,
  [\href{http://arxiv.org/abs/1407.1553}{{\tt arXiv:1407.1553}}].

\bibitem{Contopanagos:1997nh}
H.~Contopanagos, E.~Laenen, and G.~Sterman, {\it Sudakov factorization and
  resummation},  {\em Nucl. Phys.} {\bf B484} (1997) 303--330,
  [\href{http://arxiv.org/abs/hep-ph/9604313}{{\tt hep-ph/9604313}}].

\bibitem{Bonocore:2015esa}
D.~Bonocore, E.~Laenen, L.~Magnea, S.~Melville, L.~Vernazza, and C.~D. White,
  {\it {A factorization approach to next-to-leading-power threshold
  logarithms}},  {\em JHEP} {\bf 1506} (2015) 008,
  [\href{http://arxiv.org/abs/1503.05156}{{\tt arXiv:1503.05156}}].

\bibitem{Bonocore:2016awd}
D.~Bonocore, E.~Laenen, L.~Magnea, L.~Vernazza, and C.~D. White, {\it
  {Non-Abelian factorisation for next-to-leading-power threshold logarithms}},
  {\em JHEP} {\bf 1612} (2016) 121,
  [\href{http://arxiv.org/abs/1610.06842}{{\tt arXiv:1610.06842}}].

\bibitem{Gervais:2017yxv}
H.~Gervais, {\it {Soft Photon Theorem for High Energy Amplitudes in Yukawa and
  Scalar Theories}},  {\em Phys. Rev.} {\bf D95} (2017), no.~12 125009,
  [\href{http://arxiv.org/abs/1704.00806}{{\tt arXiv:1704.00806}}].

\bibitem{Gervais:2017zky}
H.~Gervais, {\it {Soft Graviton Emission at High and Low Energies in Yukawa and
  Scalar Theories}},  {\em Phys. Rev.} {\bf D96} (2017), no.~6 065007,
  [\href{http://arxiv.org/abs/1706.03453}{{\tt arXiv:1706.03453}}].

\bibitem{Gervais:2017zdb}
H.~Gervais, {\em {Soft Radiation Theorems at All Loop Order in Quantum Field
  Theory}}.
\newblock PhD thesis, SUNY, Stony Brook, 2017-08-04.

\bibitem{Larkoski:2014bxa}
A.~J. Larkoski, D.~Neill, and I.~W. Stewart, {\it {Soft Theorems from Effective
  Field Theory}},  {\em JHEP} {\bf 1506} (2015) 077,
  [\href{http://arxiv.org/abs/1412.3108}{{\tt arXiv:1412.3108}}].

\bibitem{Kolodrubetz:2016uim}
D.~W. Kolodrubetz, I.~Moult, and I.~W. Stewart, {\it {Building Blocks for
  Subleading Helicity Operators}},  {\em JHEP} {\bf 1605} (2016) 139,
  [\href{http://arxiv.org/abs/1601.02607}{{\tt arXiv:1601.02607}}].

\bibitem{Moult:2016fqy}
I.~Moult, L.~Rothen, I.~W. Stewart, F.~J. Tackmann, and H.~X. Zhu, {\it
  {Subleading Power Corrections for N-Jettiness Subtractions}},  {\em Phys.
  Rev.} {\bf D95} (2017), no.~7 074023,
  [\href{http://arxiv.org/abs/1612.00450}{{\tt arXiv:1612.00450}}].

\bibitem{Moult:2017rpl}
I.~Moult, I.~W. Stewart, and G.~Vita, {\it {A subleading operator basis and
  matching for gg $\rightarrow$ H}},  {\em JHEP} {\bf 1707} (2017) 067,
  [\href{http://arxiv.org/abs/1703.03408}{{\tt arXiv:1703.03408}}].

\bibitem{Feige:2017zci}
I.~Feige, D.~W. Kolodrubetz, I.~Moult, and I.~W. Stewart, {\it {A Complete
  Basis of Helicity Operators for Subleading Factorization}},  {\em JHEP} {\bf
  1711} (2017) 142, [\href{http://arxiv.org/abs/1703.03411}{{\tt
  arXiv:1703.03411}}].

\bibitem{Chang:2017atu}
C.-H. Chang, I.~W. Stewart, and G.~Vita, {\it {A Subleading Power Operator
  Basis for the Scalar Quark Current}},  {\em JHEP} {\bf 1804} (2018) 041,
  [\href{http://arxiv.org/abs/1712.04343}{{\tt arXiv:1712.04343}}].

\bibitem{Beneke:2004in}
M.~Beneke, F.~Campanario, T.~Mannel, and B.~D. Pecjak, {\it {Power corrections
  to anti-B $\rightarrow$ X(u) l anti-nu (X(s) gamma) decay spectra in the
  'shape-function' region}},  {\em JHEP} {\bf 0506} (2005) 071,
  [\href{http://arxiv.org/abs/hep-ph/0411395}{{\tt hep-ph/0411395}}].

\bibitem{Moult:2017jsg}
I.~Moult, L.~Rothen, I.~W. Stewart, F.~J. Tackmann, and H.~X. Zhu, {\it {N
  -jettiness subtractions for $gg\to H$ at subleading power}},  {\em Phys.
  Rev.} {\bf D97} (2018), no.~1 014013,
  [\href{http://arxiv.org/abs/1710.03227}{{\tt arXiv:1710.03227}}].

\bibitem{Ebert:2018lzn}
M.~A. Ebert, I.~Moult, I.~W. Stewart, F.~J. Tackmann, G.~Vita, and H.~X. Zhu,
  {\it {Power Corrections for N-Jettiness Subtractions at ${\cal
  O}(\alpha_s)$}},  {\em JHEP} {\bf 1812} (2018) 084,
  [\href{http://arxiv.org/abs/1807.10764}{{\tt arXiv:1807.10764}}].

\bibitem{Ebert:2018gsn}
M.~A. Ebert, I.~Moult, I.~W. Stewart, F.~J. Tackmann, G.~Vita, and H.~X. Zhu,
  {\it {Subleading power rapidity divergences and power corrections for
  q$_{T}$}},  {\em JHEP} {\bf 1904} (2019) 123,
  [\href{http://arxiv.org/abs/1812.08189}{{\tt arXiv:1812.08189}}].

\bibitem{Moult:2018jjd}
I.~Moult, I.~W. Stewart, G.~Vita, and H.~X. Zhu, {\it {First Subleading Power
  Resummation for Event Shapes}},  {\em JHEP} {\bf 1808} (2018) 013,
  [\href{http://arxiv.org/abs/1804.04665}{{\tt arXiv:1804.04665}}].

\bibitem{Beneke:2018gvs}
M.~Beneke, A.~Broggio, M.~Garny, S.~Jaskiewicz, R.~Szafron, L.~Vernazza, and
  J.~Wang, {\it {Leading-logarithmic threshold resummation of the Drell-Yan
  process at next-to-leading power}},  {\em JHEP} {\bf 1903} (2019) 043,
  [\href{http://arxiv.org/abs/1809.10631}{{\tt arXiv:1809.10631}}].

\bibitem{Bhattacharya:2018vph}
A.~Bhattacharya, I.~Moult, I.~W. Stewart, and G.~Vita, {\it {Helicity Methods
  for High Multiplicity Subleading Soft and Collinear Limits}},
  \href{http://arxiv.org/abs/1812.06950}{{\tt arXiv:1812.06950}}.

\bibitem{Moult:2019mog}
I.~Moult, I.~W. Stewart, and G.~Vita, {\it {Subleading Power Factorization with
  Radiative Functions}},  \href{http://arxiv.org/abs/1905.07411}{{\tt
  arXiv:1905.07411}}.

\bibitem{DelDuca:2017twk}
V.~Del~Duca, E.~Laenen, L.~Magnea, L.~Vernazza, and C.~D. White, {\it
  {Universality of next-to-leading power threshold effects for colourless final
  states in hadronic collisions}},  {\em JHEP} {\bf 1711} (2017) 057,
  [\href{http://arxiv.org/abs/1706.04018}{{\tt arXiv:1706.04018}}].

\bibitem{Bonocore:2014wua}
D.~Bonocore, E.~Laenen, L.~Magnea, L.~Vernazza, and C.~D. White, {\it {The
  method of regions and next-to-soft corrections in Drell-Yan production}},
  {\em Phys. Lett.} {\bf B742} (2015) 375--382,
  [\href{http://arxiv.org/abs/1410.6406}{{\tt arXiv:1410.6406}}].

\bibitem{Bahjat-Abbas:2018hpv}
N.~Bahjat-Abbas, J.~Sinninghe~Damst\'{e}, L.~Vernazza, and C.~D. White, {\it
  {On next-to-leading power threshold corrections in Drell-Yan production at
  N$^3$LO}},  {\em JHEP} {\bf 1810} (2018) 144,
  [\href{http://arxiv.org/abs/1807.09246}{{\tt arXiv:1807.09246}}].

\bibitem{Boughezal:2016zws}
R.~Boughezal, X.~Liu, and F.~Petriello, {\it {Power Corrections in the
  N-jettiness Subtraction Scheme}},  {\em JHEP} {\bf 1703} (2017) 160,
  [\href{http://arxiv.org/abs/1612.02911}{{\tt arXiv:1612.02911}}].

\bibitem{Boughezal:2018mvf}
R.~Boughezal, A.~Isgr{\`o}, and F.~Petriello, {\it {Next-to-leading-logarithmic
  power corrections for $N$-jettiness subtraction in color-singlet
  production}},  {\em Phys. Rev.} {\bf D97} (2018), no.~7 076006,
  [\href{http://arxiv.org/abs/1802.00456}{{\tt arXiv:1802.00456}}].

\bibitem{Basu:2007nu}
R.~Basu, E.~Laenen, A.~Misra, and P.~Motylinski, {\it {Soft-collinear effects
  in prompt photon production}},  {\em Phys. Rev.} {\bf D76} (2007) 014010,
  [\href{http://arxiv.org/abs/0704.3180}{{\tt arXiv:0704.3180}}].

\bibitem{Basu:2012ma}
R.~Basu, E.~Laenen, A.~Misra, and P.~Motylinski, {\it {Soft-collinear effects
  for prompt photon production via fragmentation}},
  \href{http://arxiv.org/abs/1204.2503}{{\tt arXiv:1204.2503}}.

\bibitem{Misra:2018nsy}
M.~van Beekveld, W.~Beenakker, E.~Laenen, A.~Misra, P.~Motylinski, and R.~Basu,
  {\it {Soft-Collinear Effects in Threshold and Joint Resummation}},  {\em Few
  Body Syst.} {\bf 59} (2018), no.~5 99.

\bibitem{Catani:1996yz}
S.~Catani, M.~L. Mangano, P.~Nason, and L.~Trentadue, {\it The resummation of
  soft gluons in hadronic collisions},  {\em Nucl. Phys.} {\bf B478} (1996)
  273--310, [\href{http://arxiv.org/abs/hep-ph/9604351}{{\tt hep-ph/9604351}}].

\bibitem{Vogt:2004ns}
A.~Vogt, {\it {Efficient evolution of unpolarized and polarized parton
  distributions with QCD-PEGASUS}},  {\em Comput. Phys. Commun.} {\bf 170}
  (2005) 65--92, [\href{http://arxiv.org/abs/hep-ph/0408244}{{\tt
  hep-ph/0408244}}].

\bibitem{Catani:2003zt}
S.~Catani, D.~de~Florian, M.~Grazzini, and P.~Nason, {\it Soft-gluon
  resummation for {H}iggs boson production at hadron colliders},  {\em JHEP}
  {\bf 0307} (2003) 028, [\href{http://arxiv.org/abs/hep-ph/0306211}{{\tt
  hep-ph/0306211}}].

\bibitem{Gluck:1989ze}
M.~Gluck, E.~Reya, and A.~Vogt, {\it Radiatively generated parton distributions
  for high-energy collisions},  {\em Z. Phys.} {\bf C48} (1990) 471--482.

\bibitem{Furmanski:1980cm}
W.~Furmanski and R.~Petronzio, {\it Singlet parton densities beyond leading
  order},  {\em Phys. Lett.} {\bf 97B} (1980) 437.

\bibitem{Furmanski:1982cw}
W.~Furmanski and R.~Petronzio, {\it Lepton - hadron processes beyond leading
  order in quantum chromodynamics},  {\em Zeit. Phys.} {\bf C11} (1982) 293.

\bibitem{Gluck:1993im}
M.~Gluck, E.~Reya, and A.~Vogt, {\it Comparing radiatively generated parton
  distributions with recent measurements of $f_2(x,q^2)$ in the small x
  region},  {\em Phys. Lett.} {\bf B306} (1993) 391--394.

\bibitem{Floratos:1981hs}
E.~G. Floratos, C.~Kounnas, and R.~Lacaze, {\it Higher order qcd effects in
  inclusive annihilation and deep inelastic scattering},  {\em Nucl. Phys.}
  {\bf B192} (1981) 417.

\bibitem{vanBeekveld:2019prq}
M.~van Beekveld, W.~Beenakker, E.~Laenen, and C.~D. White, {\it
  {Next-to-leading power threshold effects for inclusive and exclusive
  processes with final state jets}},
  \href{http://arxiv.org/abs/1905.08741}{{\tt arXiv:1905.08741}}.

\bibitem{Mathews:2004pu}
P.~Mathews et~al., {\it Working group report: Quantum chromodynamics},  {\em
  Pramana} {\bf 63} (2004) 1367--1379. Proceedings of the Eighth Workshop on
  High Energy Physics Phenomenology, Mumbai, India, Jan. 2004.

\bibitem{PhysRevD.45.3986}
M.~Gl\"uck, E.~Reya, and A.~Vogt, {\it Parton structure of the photon beyond
  the leading order},  {\em Phys. Rev.} {\bf D45} (Jun, 1992) 3986--3994.

\bibitem{Harland-Lang:2014zoa}
L.~A. Harland-Lang, A.~D. Martin, P.~Motylinski, and R.~S. Thorne, {\it {Parton
  distributions in the LHC era: MMHT 2014 PDFs}},  {\em Eur. Phys. J.} {\bf
  C75} (2015), no.~5 204, [\href{http://arxiv.org/abs/1412.3989}{{\tt
  arXiv:1412.3989}}].

\bibitem{Kaufmann:2017lsd}
T.~Kaufmann, A.~Mukherjee, and W.~Vogelsang, {\it {Recent developments on
  parton-to-photon fragmentation functions}},  {\em CERN Proc.} {\bf 1} (2018)
  211, [\href{http://arxiv.org/abs/1708.06683}{{\tt arXiv:1708.06683}}].

\bibitem{Catani:2018krb}
S.~Catani, L.~Cieri, D.~de~Florian, G.~Ferrera, and M.~Grazzini, {\it {Diphoton
  production at the LHC: a QCD study up to NNLO}},  {\em JHEP} {\bf 1804}
  (2018) 142, [\href{http://arxiv.org/abs/1802.02095}{{\tt arXiv:1802.02095}}].

\bibitem{Gluck:1993zx}
M.~Gluck, E.~Reya, and A.~Vogt, {\it Parton fragmentation into photons beyond
  the leading order},  {\em Phys. Rev.} {\bf D48} (1993) 116--128.

\bibitem{Laenen:2000ht}
E.~Laenen, G.~F. Sterman, and W.~Vogelsang, {\it {Combined recoil and threshold
  resummation for hard scattering cross-sections}},
  \href{http://arxiv.org/abs/hep-ph/0010184}{{\tt hep-ph/0010184}}.

\bibitem{Aversa:1988vb}
F.~Aversa, P.~Chiappetta, M.~Greco, and J.~P. Guillet, {\it Qcd corrections to
  parton-parton scattering processes},  {\em Nucl. Phys.} {\bf B327} (1989)
  105.

\bibitem{Aurenche:1983ws}
P.~Aurenche, A.~Douiri, R.~Baier, M.~Fontannaz, and D.~Schiff, {\it Prompt
  photon production at large p(t) in qcd beyond the leading order},  {\em Phys.
  Lett.} {\bf B140} (1984) 87.

\bibitem{Aurenche:1987fs}
P.~Aurenche, R.~Baier, M.~Fontannaz, and D.~Schiff, {\it Prompt photon
  production at large p(t) scheme invariant qcd predictions and comparison with
  experiment},  {\em Nucl. Phys.} {\bf B297} (1988) 661.

\bibitem{Gordon:1993qc}
L.~E. Gordon and W.~Vogelsang, {\it Polarized and unpolarized prompt photon
  production beyond the leading order},  {\em Phys. Rev.} {\bf D48} (1993)
  3136--3159.

\end{thebibliography}\endgroup

\end{document}